\documentclass{aa}
\usepackage{graphics,epsfig}
\usepackage{verbatim}
\usepackage{amssymb}
\usepackage{natbib}[1999/03/31]
\bibpunct{(}{)}{;}{a}{}{,}

\begin{document}
\thesaurus{03(11.06.1; 11.09.2; 11.16.1; 11.05.2) }

\title{Tidal dwarf candidates in a sample of interacting
  galaxies\thanks{Based on observations collected at the European
    Southern Observatory, La Silla, Chile (ESO No
    058.A-0260).}$^,$\thanks{Tables 5-14 are only available in electronic
    form at the CDS via anonymous ftp to 130.79.128.5 or via
    http://cds.u-strasbg.fr/Abstract.html}
}
\author{Peter M.~Weilbacher \inst{1} \and P.-A.~Duc \inst{2,3} \and
  U.~Fritze-v.Alvensleben \inst{1} \and P.~Martin \inst{4} \and
  K.J.~Fricke \inst{1} }
\institute{Universit\"atssternwarte, Geismarlandstr.~11, D-37083
  G\"ottingen, Germany 
  \and CNRS URA 2052 and CEA, DSM, DAPNIA, Service d'Astrophysique,
  Centre d'Etudes de Saclay, 91191 Gif-sur-Yvette Cedex, France
  \and Institute of Astronomy, University of Cambridge, Madingley
  Road, Cambridge, CB3 0HA, UK
  \and Canada-France-Hawaii Telescope, P.O.~Box 1597, Kamuela, HI
  96743, USA }
\authorrunning{P.~M.~Weilbacher et al.}  
\titlerunning{TDG Candidates in Interacting Systems}
\offprints{P.~Weilbacher, \\
  weilbach@uni-sw.gwdg.de}
\date{Received 30 July 1999 / Accepted 28 March 2000}
\maketitle

\begin{abstract}
  We present deep optical $B$,$V\!$,$R$ images of a sample of 10
  interacting systems which were selected for their resemblance to
  disturbed galaxies at high redshift.  Photometry is performed on
  knots in the tidal features of the galaxies. We calculate a grid of
  evolutionary synthesis models with two metallicities and various
  burst strengths for systems consisting of some fraction of the
  stellar population of a progenitor spiral plus starburst.  By
  comparison with two-color diagrams we interpret the photometric
  data, select from a total of about 100 condensations 36 star-forming
  objects that are located in the tidal features and predict their
  further evolution. Being more luminous by 4 mag than normal
  \ion{H}{ii} regions we argue that these objects could be tidal dwarf
  galaxies or their progenitors, although they differ in number and
  mean luminosity from the already known tidal dwarf galaxies
  typically located at the end of tidal tails in nearby giant
  interacting systems. From comparison with our models we note that
  all objects show young burst ages. The young stellar component
  formed in these tidal dwarf candidates contributes up to 18\% to the
  total stellar mass at the end of the starburst and dominates the
  optical luminosity. This may result in fading by up to 2.5 mag in
  $B$ during the next 200 Myrs after the burst.
  
  \keywords{Galaxies: formation -- Galaxies: interactions -- Galaxies:
    photometry -- Galaxies: evolution}
\end{abstract}

\section{Introduction}\label{Intro-Sec}
Most investigations of interacting and merging galaxies have
essentially focussed on the phenomena occuring in their inner regions.
However, for a few spectacular cases like the Antennae \citep{MLM91}, Arp 105
\citep{DM94,DBW97} and NGC 7252 \citep{HM95} the enormous tidal tails
were investigated in some detail. Massive condensations of stars
and \ion{H}{i} were found in these tails, the so called Tidal Dwarf
Galaxies (TDGs).

TDGs are characterized by a luminosity comparable with that of typical
dwarf galaxies, but which span a range in oxygen abundance of $8.3
\lesssim 12+\log(O/H) \lesssim 8.6$ equivalent to metallicities of
$4\cdot10^{-3} \lesssim Z_\mathrm{TDG} \lesssim 8\cdot10^{-3}$ with a
mean of $Z \approx 7\cdot10^{-3} \approx Z_{\sun} / 2.6$.\footnote{For
  metallicity we use the notation of $Z$ (as the proportion of heavy
  elements given in the Geneva stellar tracks which our models are
  based on) throughout this paper and give a comparison with the solar
  value $Z_{\sun} = 0.018$ where appropriate.} Metallicities of TDGs
are therefore higher than those of other dwarf galaxies with
comparable luminosity \citep{DM97Mess,DM98}.  TDGs have blue colors as
a result of an active starburst. Most known TDGs have high \ion{H}{i}
masses from which total masses of the order of 10$^9$ M$_{\sun}$ have
been estimated.  They are gravitationally bound and in a few
interacting systems evidence for their kinematical independence has
been found \citep[e.g.][]{DBW97,DM98}.  These systems together with
the strong increase of the galaxy merger rate with redshift gave rise
to speculations about a possible contribution of TDGs to the faint
blue galaxy excess at high redshift \citep{FvAM98}. \citet{BH92}
observed the formation of TDGs in their numerical simulations. Their
3D N-body/SPH-code created several massive bound condensations along
the tidal tails. At first, their mass is dominated by the stellar
component.  Later on, gas from the tail might fall into these
condensations.  \citet{EKT93} proposed a different scenario.  They
have used a 2D N-body code including dissipation for the gas component
in which gaseous condensations form first.

Observations in most cases show both components, neutral gas and
stars, along the tails as well as in their condensations. In an
optically indentified condensation, determination of the relative
fraction of young stars -- born in situ from the collapse of tidally
extracted \ion{H}{i} -- to old stars -- pulled out from the progenitor
disk(s) -- might constrain the above mentioned numerical models.
Because the systems studied here are too distant to be resolved into
stars, estimating these relative stellar fractions can only be done
from multi-band integrated photometric measurements in comparison with
an evolutionary synthesis model.  Photometric work on the tidal
features of several classical interacting systems has been carried out
by \citet{SWS90}.  Studies by \citet{AM97} and \citet{BAM98}
concentrated on ring galaxies.  \citet{DMT98} have systematically
catalogued all faint non-stellar objects around disturbed galaxies.
Part of them might be TDG candidates.

These previous investigations have focused on disk-disk interacting
systems with a well-understood morphology, e.g. well defined tails.
However locally the majority of disturbed galaxy systems, for instance
in the Catalogue of Southern Peculiar Galaxies \citep{AM87}, do not
have long tails.  In particular disk-spheroid collisions and 3-body
systems leading to chaotic morphologies are quite common.  HST deep
surveys with the WFPC2 show that the percentage of disturbed
(i.e.~interacting, merging or peculiar) galaxies has been much higher
in the early universe than today \citep{vdB96}.  NICMOS observations
in the near-infrared have shown that these disturbances are not only
due to irregularities in the distribution of star forming regions, but
reflect real morphological features of interactions \citep{CBD98}.
These galaxies at high redshift often do not show long tidal tails
either because they are absent or because they have too low a surface
brightness. Examining a local sample which resembles these high
redshift interacting galaxies is therefore important to learn about
their evolution.

In this paper we extend previous investigations of a few individual
systems to deep optical photometry of a sample of peculiar galaxies.
The objects from our sample mostly do not possess prominent tidal
tails, but definitely show signs of merging, as e.g.~multiple nuclei
or disturbed morphology and nearby companions.  The sample is
extracted from the Catalogue of Southern Peculiar Galaxies
\citep{AM87}.  It includes disk-disk, disk-elliptical, dwarf-dwarf and
3-body systems observed at different stages of the interaction.

Spectrophotometric evolutionary synthesis models have been used for
various applications of normal galaxy evolution. More recently
starburst events were modeled as well. These models assume that the
standard star-formation rate (SFR) of an underlying galaxy is
increased to simulate a starburst \citep{FvAG94a,LH95}. This technique
has been applied to reproduce the star formation history of various
classes of objects from merging galaxies \citep{FvAG94b} to blue
compact galaxies \citep{KFL95}.

We take an approach similar to the latter and compute a grid of models
specifically aimed to describe the properties found in previous
investigations of TDGs.  We do not use solar-abundance tracks in our
models but instead two more realistic metallicities in the range
expected for TDGs ($Z = 1\cdot10^{-3}, 8\cdot10^{-3}$). As a primary
goal, these models are used to select candidate TDGs or TDG
progenitors.  Based on their colors only, background objects can be
excluded not only with statistical methods but also one by one if they
disagree with any kind of TDG model.  Furthermore, this technique
allows to derive the present ratios of old to young stellar mass in
these TDG candidates, to predict its evolution and hence to constrain
the above-mentioned formation scenarios of TDGs.  Obviously
spectrophotometric data is required to confirm the photometric
results.

The structure of the paper is as follows. First we present our data
reduction and analysis techniques (Sect.~\ref{ObsData-Sec}). We
describe in Sect.~\ref{Model-Sec} the evolutionary synthesis models
and the parameters used to interpret the photometric data.  In
Sect.~\ref{Results-Sec} we present our sample, compare with the model
grid and comment on individual features of the interacting systems.
In Sect.~\ref{Disc-Sec} we discuss more general results and in
particular compare the tidal objects identified in our sample with
those produced in numerical models and with the few typical TDGs
studied so far.

\section{Observations and data reduction}\label{ObsData-Sec}
\subsection{Data and Calibration}
Optical $B$,$V\!$,$R$ images of 10 galaxies were obtained in visitor
mode on April 10th, 1996 and in service mode from February to April
1997 using SUSI on the NTT. The SUSI camera has a field of view of
2\farcm2$\times$2\farcm2 and a pixel scale of 0\farcs13.  The objects
of our sample with observing dates, seeing conditions and total
exposure times are given in Table~\ref{DataTab}.

\begin{table*}
\caption{Observation log}
\label{DataTab}
\begin{tabular}{r c c c rrr}
\hline 
\multicolumn{1}{c}{Field} 
    & Object      & Observing Date            & Seeing & \multicolumn{3}{c}{total Exposure Time [s]} \\
\multicolumn{1}{c}{No.}  
    &             &                           &    [$''$]  &  \multicolumn{1}{c}{\hspace*{5pt}$B$}   &   \multicolumn{1}{c}{\hspace*{7pt}$V$}   &   \multicolumn{1}{c}{$R$}               \\
\hline 
 1  & AM 0529-565 & 02./03.03.97              & 0.8 &  \hspace*{2pt} 2400  & \hspace*{4pt} 1800 & 1800                   \\
 2  & AM 0537-292 & 28.02.97                  & 0.9 &  1200  &   900 &  900                   \\
 3  & AM 0547-244 & 01.03.97                  & 1.1 &  1200  &   900 &  900                   \\
 4  & AM 0607-444 & 03.03.97                  & 0.8 &  1200  &   900 &  900                   \\
 5  & AM 0642-645 & 08.04.97                  & 1.3 &  1200  &   900 &  900                   \\
 6  & AM 0748-665 & 16.03./03.04.97, 10.04.96 & 1.1 &  3600  &  3000 & 3000                   \\
 7  & AM 1054-325 & 11.04.                    & 0.8 &  1200  &   900 &  900                   \\
 8  & AM 1208-273 & 03.04.97                  & 1.0 &  1200  &   900 &  900                   \\
 9  & AM 1325-292 & 03.04.97                  & 0.9 &  1200  &   900 &  900                   \\
10  & AM 1353-272 & 15./16.03.97              & 0.8 &  2400  &  1200 & 1200                   \\
\hline
\end{tabular} \\
\end{table*}

The data reduction was performed in IRAF using the {\sc ccdred}
package with the classical procedures of bias-subtraction and
flat-fielding. In order to correct the sky background for some
residuals we constructed illumination frames by modeling the
background fitted at positions not contaminated by object light.
Cosmic ray hits were removed using {\sc cosmicrays}.  The images in
all filters were then registered and PSF-matched using several
reference stars.  Foreground stars were, when possible, deleted and
replaced by a background fit using the {\sc imedit} task.

\begin{figure}
        \resizebox{\hsize}{!}{\includegraphics{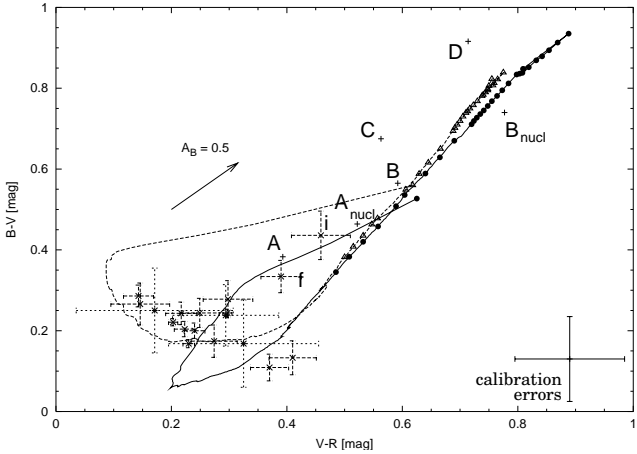}}
        \caption{Plot of the objects in field 10 of AM 1353-272. The
          models used are 4($Z_1$) (dashed, $b=0.02$) and 6($Z_3$)
          (solid, $b=0.18$).}
        \label{AM1353-272_plot}
\end{figure}

\begin{figure}
        \resizebox{\hsize}{!}{\includegraphics{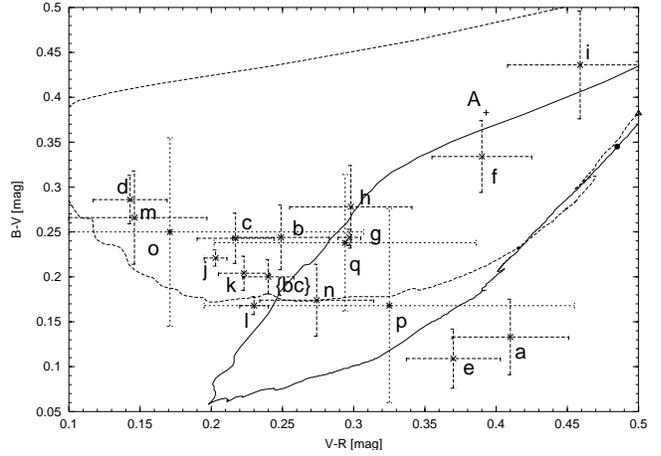}}
        \caption{Close-up of Fig.~\ref{AM1353-272_plot}. The knots are 
          now labeled.}
        \label{AM1353-272_plotZoom}
\end{figure}

Standard stars from the fields of \citet{Lan92} have been used to
derive the zeropoints and estimate the color terms.  The extinction
coefficients were taken from the La Silla extinction database provided
by the Geneva group \citep[e.g.][]{BRB95}.  With this data we could
derive a photometric calibration accurate to about 0.04 mag for most
targets. One should note that most of the observations were obtained
during the refurbishment of the NTT and that the standard stars have
not always been observed during the same night as the science targets.
Therefore additional systematic calibration errors cannot be excluded;
however, our zeropoints and color terms appear to be very stable from
one night to the other.  Their variations are smaller than 0.01 mag
i.e.~smaller than the given errors.  Except for a few cases for which
the extinction given by the Geneva group has large uncertainties, high
calibration errors might be present (e.g.~AM 0607-444).
Figs.~\ref{AM0529-565_plot} to \ref{AM1353-272_plotZoom} in
Sect.~\ref{Results-Sec} show 1$\sigma$-errorbars for the systematic
error for each field due to uncertain calibration and individual
errors for each object due to photon statistics and background noise.

We also obtained longslit-spectra of AM 0529-565, AM 1325-292, and AM
1353-272 using EMMI at the NTT in May 1998. The reduction was
performed in the standard way and the spectra were used to measure the
redshift. Further details will be given in a forthcoming paper.

\subsection{Aperture photometry}
As the features of interest within the tidal debris are irregular in
shape and located in regions with high and irregular background,
traditional photometry with circular apertures was not appropriate
since it would have included too much flux from the surrounding tidal
features. We therefore developed a reproducible technique to analyze
these knots using polygonal apertures and used circular apertures only
for objects far away from the main body of the interacting system.

The $V\!$-band intensity scale of each image was transformed to
calibrated surface brightness and contour plots were created of the
interesting regions of each frame. The objects to measure were
selected from the most prominent peaks in surface brightness.
Polygonal apertures were then defined following the faintest contour
which still allows separating the knot from surrounding tidal
features. The surface brightness level $\mu_V$ of the aperture was
therefore different for each knot and is given in Tables
5 to 14 \footnote{Tables 5 to 14 are available only in
  electronic form.}.  The sky flux was individually measured at
several positions nearby each knot, carefully avoiding the tidal
features, and averaged. We are interested in both the young and the
old stellar population which the (stellar) tail consists of, because
we explicitly model the old population in our evolutionary synthesis
models. Hence we only subtract the sky background but not the tidal
tail from the object flux. The photometry was finally performed using
the IRAF {\sc polyphot} task with the same polygonal aperture for all
photometric bands.  Tables 5 to 14 give the complete photometric results. Column 1
gives the identification, column 2 the surface brightness $\mu_V$ of
the polygonal aperture.  Columns 3 to 5 show the apparent magnitudes
in $B$, $V\!$ and $R$, while columns 6 to 8 present the optical colors
$B-V\!$, $V-R$ and $B-R$. Columns 9 to 11 show the photometric errors
of the measurements in $B$, $V\!$ and $R$. All given magnitudes and
colors are corrected for galactic extinction, using the $A_B$ values
of \citet{BH82} and extrapolating to $V\!$ and $R$ with the ratios of
\citet{SM79}.

\section{Model}\label{Model-Sec}
\subsection{Modelling the old population}
Spectrophotometric and chemical evolutionary synthesis models are used
to interpret the photometric data.  Our code is based on the work of
\citet{Krue92} which was originally used to model BCDGs \citep{KFL95}.
To include the underlying old component in our model, it starts with a
gas cloud of primordial metallicity and follows the evolution of ISM
abundances and spectrophotometric properties of the stellar population
of an undisturbed galaxy until the starburst induced by the
interaction occurs. The model includes the two basic parameters, star
formation rate (SFR) $\psi(t)$ and initial mass function (IMF).  We
use SFRs proportional to the gas to total mass ratio for Sb and Sc
models and a constant SFR for Sd with efficiencies appropriate for the
respective galaxy types \citep[see e.g.][]{FvAG94a}.

As our observations cannot constrain the IMF, we use the IMF from
\citet{Sca86} in the mass range 0.15 to 120 M$_\odot$ and, for
simplicity, assume the same IMF for the progenitor spiral and the
starburst. To cover a Hubble time of evolution for the undisturbed
progenitor spiral and have a good time resolution during and after the
burst, we use a variable timestep as outlined by \citet{Krue92}.

\subsection{Input physics}
The model uses Geneva stellar evolutionary tracks
\citep{SSM92,CMM93,SCM93,SMM93,CMM96}. We present models for two
different metallicities $Z_1 = 1\cdot10^{-3} \approx Z_{\sun} / 18$
and $Z_3 = 8 \cdot 10^{-3} \approx Z_{\sun} / 2.3$.

As shown by \citet{KFL95} the broadband colors are dominated by
nebular line and continuum emission in the early phases of a strong
burst. We therefore include nebular emission in our models using the
compilation from \citet{Sdk97} to calculate the number of Lyman
continuum photons emitted per second $N_\mathrm{Lyc}$ of our stellar
population as a function of time. With the standard formula
\[
  F(\mathrm{H}_{\beta}) = 4.757\cdot 10^{-13} 
                          f N_\mathrm{Lyc} \textrm{ erg s}^{-1}
\]
we obtain the H$_{\beta}$-flux. Fluxes of the other hydrogen lines are
computed from their line ratios relative to H$_{\beta}$. $f$ is the
proportion of Lyman continuum photons not absorbed in dust. We take
$f=1.0$ for $Z_1$ and $f=0.7$ for $Z_3$.

To include other typical lines of star-forming regions we use the
observed line ratios of 27 lines in the wavelength range from 1335 to
10330 \AA~from \citet{ITL94} for low metallicity ($Z_1$) and
theoretical line ratios from \citet{Sta84} for medium metallicity
($Z_3$) models.

We chose to use the metallicities $Z_1$ and $Z_3$ given
above\footnote{Another possible metallicity given in the Geneva
  stellar tracks ($Z_2 = 4\cdot 10^{-3}$) is not used in our models
  because no emission line ratios are available for it.}.  In the
context of comparing our models to observed objects these
metallicities should not be regarded as fixed values, but more as
metallicity ranges, $Z_1$ as $0.001 < Z < 0.006$ and $Z_3$ as $0.006 <
Z < 0.012$.

\subsection{TDG models}
To model both the star-forming young and the old component of a TDG
with our 1-zone model, we let a typical part of a disk galaxy evolve
with its appropriate SFR over 13 Gyrs. After 13 Gyrs a starburst is
assumed to occur (triggered by an interaction) in this part of the
model galaxy assumed to form a TDG.  In the burst the SFR is set to a
maximum value $\psi_0$ and decreases exponentially with a timescale
$\tau_\mathrm{B}$.

The time of the starburst is inferred from a rough age of present-day
galaxies of 14 Gyr in a reasonable cosmological model.  We look back
roughly 1 Gyr, the assumed epoch for the starburst following an
interaction. The model is followed for another 3 Gyrs after the burst.
The galaxy age at the onset of the burst does not significantly change
the color evolution during or after the burst.

\begin{table}
\caption{Set of model parameters.}
  \label{ModelGrid-Tab}
\begin{tabular}{r crc cc}
\hline
Run & $\tau_\mathrm{B}$ & \multicolumn{1}{c}{$\psi_0$} 
                                        & $b$ & \multicolumn{2}{c}{$L_\mathrm{young}/L_\mathrm{tot}$} \\
 ID &     [yr]       & [M$_{\sun}$ yr$^{-1}$] & &  $B$ & $R$ \\
\hline
1 &     $5\cdot 10^5$       &      100 & 0.01 & 0.56 & 0.37 \\
2 &     $5\cdot 10^5$       &      500 & 0.05 & 0.87 & 0.75 \\
3 &     $5\cdot 10^5$       &     1000 & 0.10 & 0.93 & 0.85 \\
\hline
4 &     $1\cdot 10^6$       &      100 & 0.02 & 0.69 & 0.49 \\
5 &     $1\cdot 10^6$       &      500 & 0.09 & 0.92 & 0.83 \\
6 &     $1\cdot 10^6$       &     1000 & 0.18 & 0.96 & 0.91 \\
\hline
7 &     $5\cdot 10^6$       &       50 & 0.04 & 0.76 & 0.58 \\
\hline
\end{tabular}
\end{table}

According to previous observations the model with metallicity of $Z_3$
should a priori best apply to TDGs.  In the dynamical formation
scenarios of \citet{BH92} and \citet{EKT93}, TDGs are predicted to
form out of material from the outer parts of the progenitor or from
the progenitor galaxy disks. This material in nearby galaxies has a
metallicity of $Z \approx 7\cdot10^{-3} \approx Z_3$
\citep{ZKH94,FGW98}.  Indeed, spectroscopy of TDGs confirms this mean
metallicity \citep{DM97Mess,DM98}.  However, the precise metallicity
expected for a particular condensation depends on the type and
luminosity of the progenitor galaxy -- late type and low luminosity
spirals being more metal poor on average than earlier type or higher
luminosity spirals -- {\it and} on the region of the parent galaxies
where this particular material is torn out from, since galaxies tend
to have negative abundance gradients in stars and the gas with
increasing radius.  Both at higher redshift in general and in the
local universe in those cases where dwarf or very low metallicity
galaxies are involved in the encounter, TDGs with lower metallicity
might also form.  However, the recycled nature of TDGs precludes
metallicities lower than about 1/20 solar.  Models with $Z_1$ predict
that they should populate distinctly different regions in optical
color-color diagrams, i.e.~have very blue $V-R$ at moderately blue
$B-V\!$.  Models with $Z_1$ have been applied in those cases.

To produce the various proportions of old and young populations
observed in TDGs, the maximum SFR and the timescale of the burst
$\tau_\mathrm{B}$ are varied. The resulting burst strength $b$ is
computed as the ratio of the mass of stars formed during the burst and
the total mass of stars ever formed in the galaxy. Our model grid is
presented in Table \ref{ModelGrid-Tab}.  Column 1 gives a code for
each model, which is used in the plots in Sect.~\ref{Results-Sec}
together with the metallicity.  Columns 2 and 3 give the max.~SFR and
decay time of the burst and column 4 the burst strength.  Columns 5
and 6 give the relative contributions of the starburst component to
the luminosities in the $B$ and $R$ bands.

\subsection{Model application and limitations}
\begin{figure}[tbp]
  \resizebox{\hsize}{!}{\includegraphics{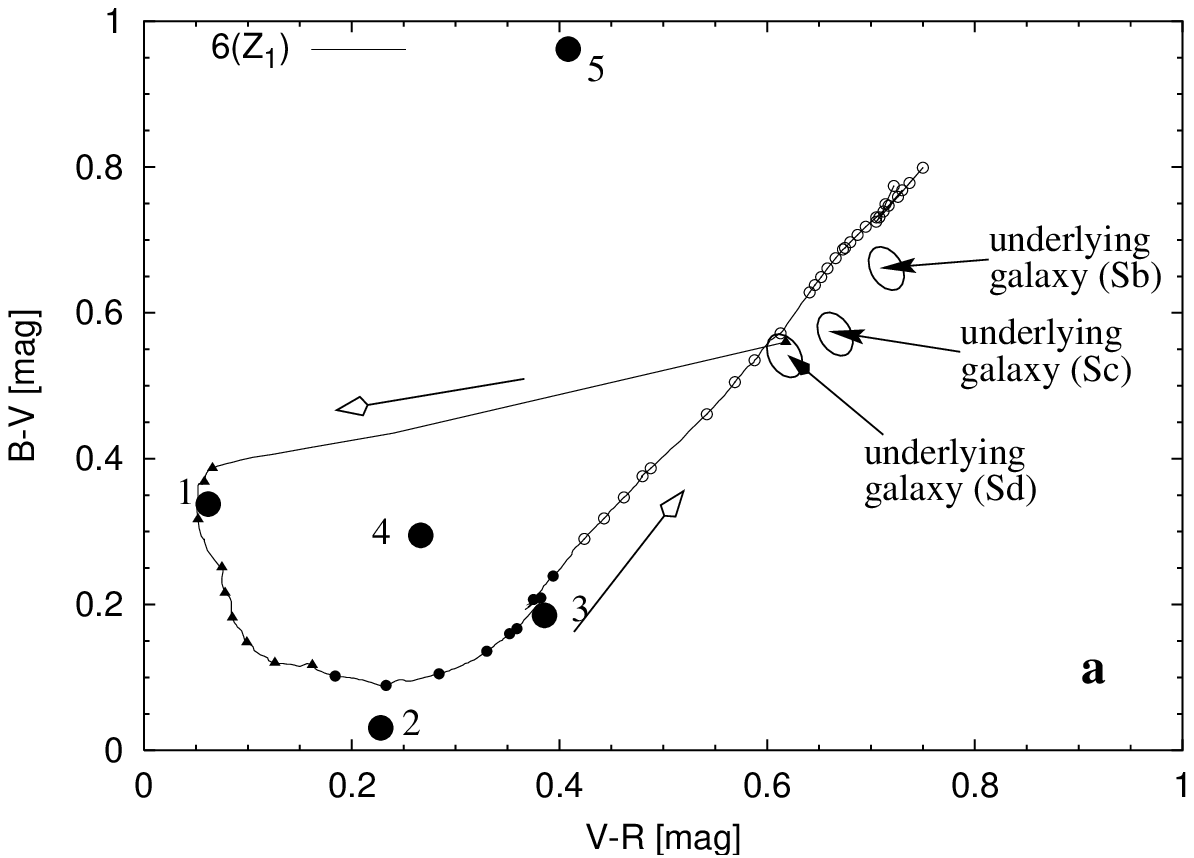}}
  \resizebox{\hsize}{!}{\includegraphics{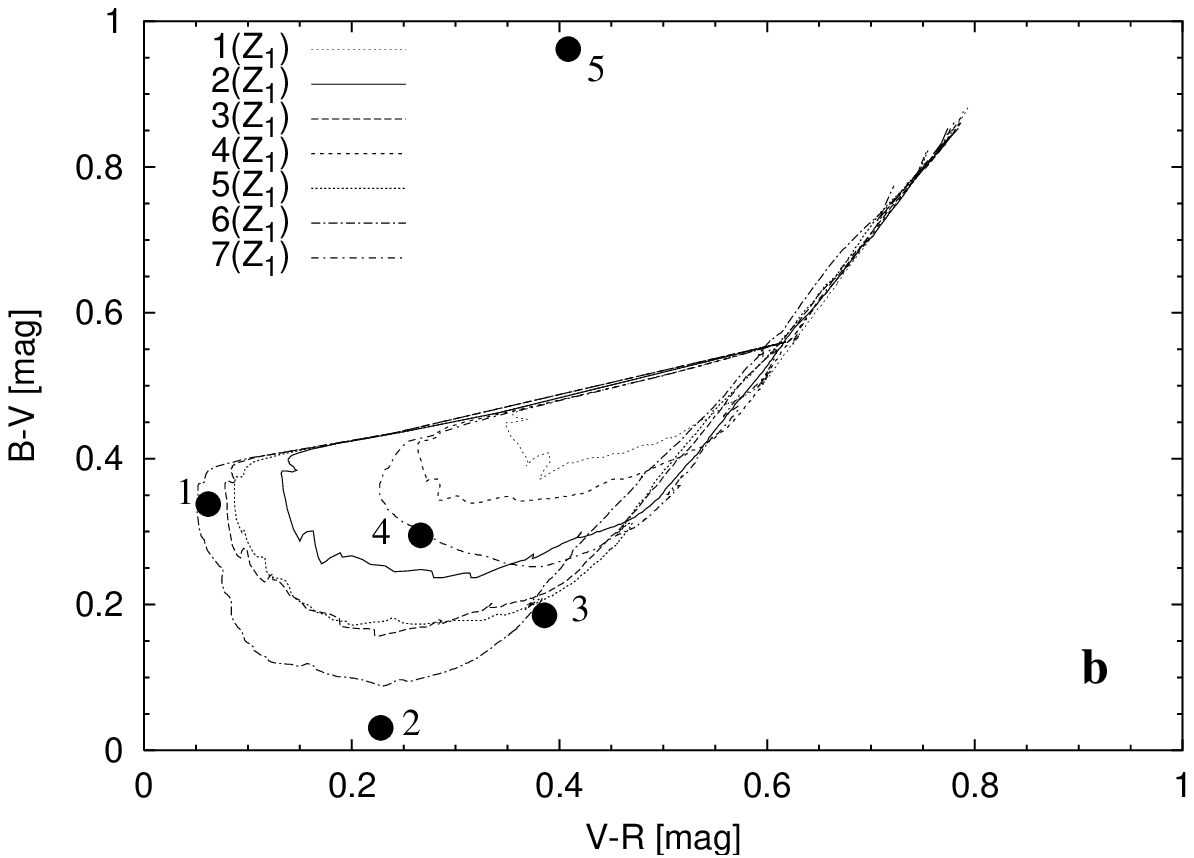}}
  \resizebox{\hsize}{!}{\includegraphics{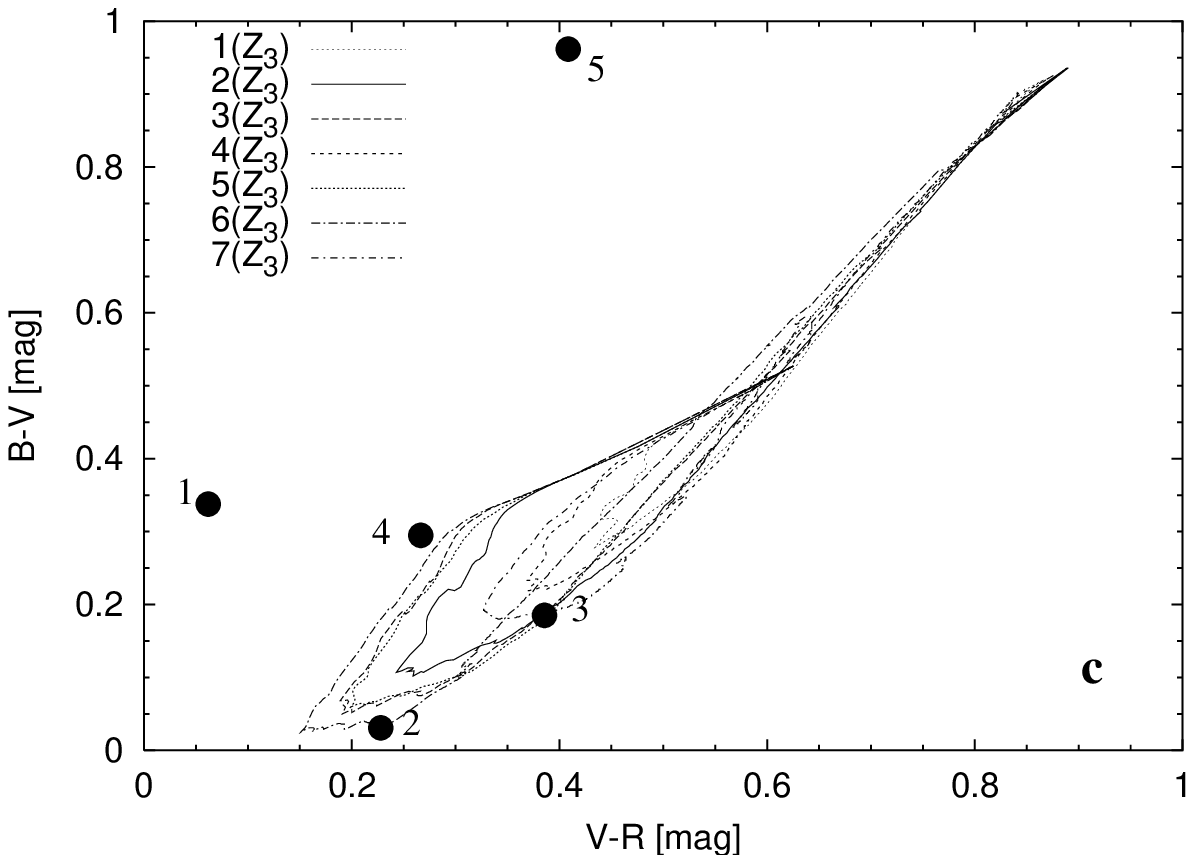}}
  \caption{Examples of model curves in a $B-V\!$ vs.~$V-R$ diagram
    for various burst parameters. Five ``data points'' are shown
    as examples, see discussion in the text. {\bf a} shows the
    strongest low metallicity burst model with tick marks indicating
    the burst age: The triangles are separated by 1 Myr starting at
    the onset of the burst, the filled circles are separated by 10
    Myrs, and the open circles by 100 Myrs. {\bf b} shows all low
    metallicity ($Z_1$) burst loops and {\bf c} all medium metallicity
    bursts.}
  \label{Example-Plot}
\end{figure}

We discuss here the application and limitations of our grid of models
to derive the properties of the observed condensations in the tidal
features.  Fig.~\ref{Example-Plot}{\bf a} details the track in a
$B-V\!$ vs.~$V-R$ two-color diagram for one given burst-model and
various assumptions for the underlying parent galaxy. The color
evolution is plotted from the start of the burst at 13 Gyrs
(underlying galaxy Sd) when a subpopulation of the progenitor spiral
is expelled into a tidal tail and begins to form a TDG. With the
burst, the colors become bluer, at first most noticeable in $V-R$ and
then in $B-V\!$ when the starburst begins to fade.  In the two-color
diagram this color evolution produces a loop which is followed
counterclockwise.  While the color evolution is very rapid at
the beginning of the burst, it slows down considerably as the burst
ages. The type of the underlying galaxy and hence its star formation
history slightly affect the starting point of the loop but do not
significantly change its shape. Due to lack of detailed knowledge of
the spectral type of the undisturbed progenitor we choose to uniformly
model the old population with a constant SFR.

Figs.~\ref{Example-Plot}{\bf b} and {\bf c} display the curves of all
models that have been probed for the two metallicities $Z_1$ and
$Z_3$, respectively. It can be seen that a higher initial burst SFR
$\psi_0$ and hence a higher burst strength $b$ produces a larger loop.
The different shapes of the loops for models with equal strength but
different metallicities is mainly caused by the variing fluxes of the
emission lines which strongly contribute to the optical broad band
colors during the burst \citep{Krue92}.

The dots plotted in Fig.~\ref{Example-Plot} show five representative
artificial data points chosen to demonstrate in howfar the various
properties of a burst can be disentangled using two colors only.  Data
point ``1'' lies in a region in the two-color plane which can only be
reached by a low metallicity model with high burst strength ($b >
0.1$) and young burst age (about 2.5 Myrs).  Data point ``2'' needs a
high burst strength ($b > 0.1$) and medium burst age (around 20 Myrs),
whatever the metallicity, the medium metallicity model providing a
better match.  Data point ``3'' is more ambiguous with respect to the
metallicity but clearly needs a much higher age (above 80 Myrs) and a
medium to strong burst ($b > 0.05$).  Depending on the metallicity
data point ``4'' can be either an intermediate strength burst (low
$Z$, $0.02 < b < 0.05$, young age $\approx 1$ Myr) or strong burst
(for medium $Z$, $b > 0.10$, age $\approx 7$ Myrs).  Additional colors
or spectroscopy are needed to resolve the ambiguity.  Data point ``5''
does not agree at all with any of our models.  Its colors are
indicative of a redshifted background galaxy (see
Sect.~\ref{BgCont-Sect} and M\"oller et al.~in prep.).

In Sect.~\ref{Results-Sec} we compare the photometry of the
star-forming knots lying in or close to the tidal features of our
interacting galaxies with predictions of various TDG models. The
primary goal is to isolate objects consistent with a TDG scenario from
background sources. A metallicity of $Z_3$ as observed for TDGs was a
priori selected unless other indicators (a metallicity measurement
from spectroscopic data, the lumosity class of the parent galaxy, or
colors that clearly indicate a low metallicity model like data point
``1'') favor a $Z_1$ model. The burst parameters were determined by
visual comparison of the position of the data points to the model
curves. The model with the smallest deviation from an observed data
point was selected as ``most likely'' model from our grid. We can also
assess which other models represent a reasonable match within the
observational uncertainties and from that derive likely ranges for the
burst age and strength of any tidal object (see Table
\ref{AllTDGCand-Tab}).  Given the individual and systematic
photometric errors and the two-dimensional nature of the data, a
numerical fitting routine did not seem to be warranted.

As the modelling of the undisturbed galaxy is created for an average
late-type galaxy the colors of the nuclei or the parent galaxies
plotted in Sect.~\ref{Results-Sec} may have strongly differing colors
(as in the case of AM 1325-292, see Fig.~\ref{AM1325-292_plot}). This
does not affect the validity of our models with respect to the TDG
candidates.

It should be noted that the comparison of observation with models only
relies on the two colors $B-V\!$ and $V-R$. To account for the effect
of dust obscuration on the colors, a larger set of photometric data
including the NIR regime or spectral information would be needed. In
the following, we neglect any internal extinction.  The effect of $A_B
= 0.5$ mag absorption is indicated by an arrow in the plots in
Sect.~\ref{Results-Sec}.  A significant extinction (of e.g.~$A_B= 0.5$
mag or more) would therefore result in underestimating the burst
strength, but would not change the final conclusions drawn from the
model.

\section{Results}\label{Results-Sec}
Figures \ref{AM0529-565} to \ref{AM1353-272} present surface
brightness images of all the systems in our sample and the two-color
diagrams for comparison with the evolutionary synthesis models. The
objects of interest are identified on the charts and the plots. The
principal objects -- those that most probably pre-existed the
interaction -- have been labeled with capital letters, while the
smaller features -- those that are TDG candidates -- have small
letters. An overview of the properties of all parent galaxies is given
in Table~\ref{Parents-Tab}. The properties of the objects for which we
have infered a tidal origin and are good candidates for TDGs or their
progenitors are listed in Table~\ref{AllTDGCand-Tab}. The model curves
shown on the diagrams are those representing a maximum number of data
points.

\begin{table}
  \caption{Properties of the parent galaxies.}
  \label{Parents-Tab}
  \begin{tabular}{l rccl}
        \hline
  Name          &\multicolumn{1}{c}{\hspace*{-10pt}Velocity}& $B$ &$M_B\,^1$& Morphology \\
                &\multicolumn{1}{c}{\hspace*{-5pt}[km s$^{-1}$]}&[mag]&[mag]&            \\
        \hline
  AM 0529-565 A &   4420$^5$  &15.83&-18.3& Merger \\
  AM 0529-565 D &   4645$^5$  &18.46&-15.4& dI \\
  AM 0537-292 A &   --~~~~    &14.97& --  & Sb?$^2$ \\
  AM 0537-292 B &   --~~~~    &16.56& --  & E \\
  AM 0547-244 A &  13164$^3$  &15.62&-20.6& S pec \\
  AM 0547-244 B &   --~~~~    &17.10& --  & S? \\
  AM 0607-444 A &   --~~~~    &15.36& --  & Sc:pec \\
  AM 0642-645 A &   --~~~~    &16.44& --  & Pec \\
  AM 0748-665 A &   --~~~~    &18.01& --  & S \\
  AM 0748-665 B &   --~~~~    &16.37& --  & E \\
  AM 1054-325 A &   3795$^4$  &14.55&-19.0& Sm \\
  AM 1054-325 B &   --~~~~    &15.31& --  & S pec \\
  AM 1208-273 A &  14778$^3$  &15.62&-20.9& Sc$^2$ \\
  AM 1208-273 B &  12433$^3$  &16.28&-19.8& Pec \\
  AM 1325-292 A &   4598$^5$  &13.90&-19.9& SB(s)b$^2$ \\
  AM 1325-292 B &   4431$^5$  &13.43&-20.4& E1 pec$^2$ \\
  AM 1353-272 A &  11791$^5$  &15.77&-20.3& Sc:pec \\
  AM 1353-272 B &  12145$^5$  &17.96&-18.1& S pec \\
        \hline
  \end{tabular}
$^1$We use $H_0 = 75$ km s$^{-1}$ throughout this paper \\
$^2$The morphological type was taken from NED \\
$^3$from \citet{DP97} \\
$^4$from \citet{SW93} \\
$^5$from Weilbacher et al.~2000 in prep. \\
\end{table}

\subsection{\object{AM 0529-565}}
\begin{figure}
        \resizebox{\hsize}{!}{\includegraphics{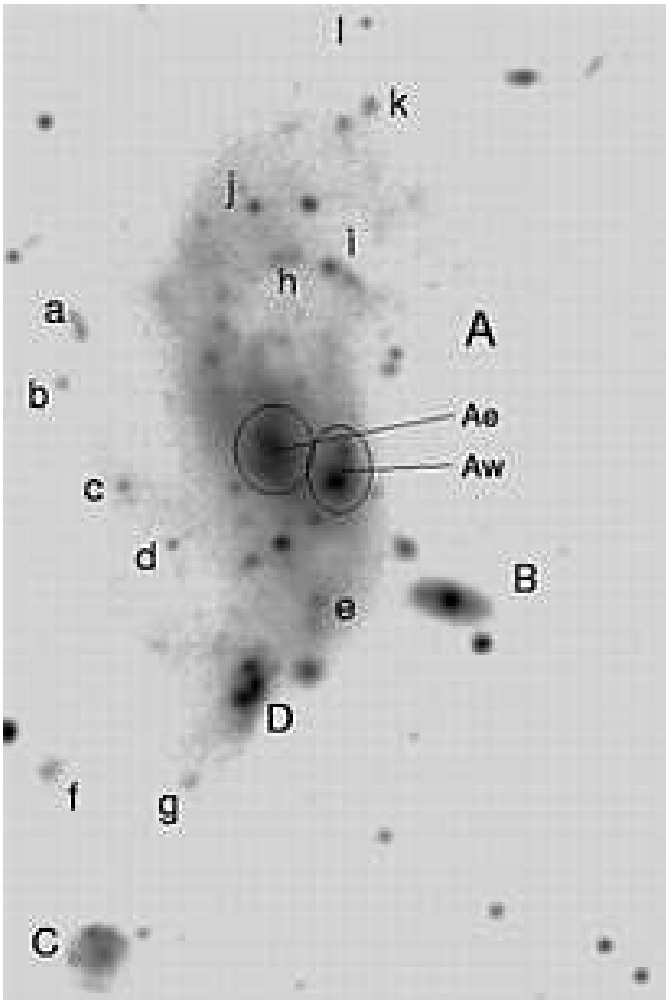}}
        \caption{Identification chart of field 1 around AM 0529-565.}
        \label{AM0529-565}
\end{figure}
\psfull 
\begin{figure}
        \resizebox{\hsize}{!}{\includegraphics{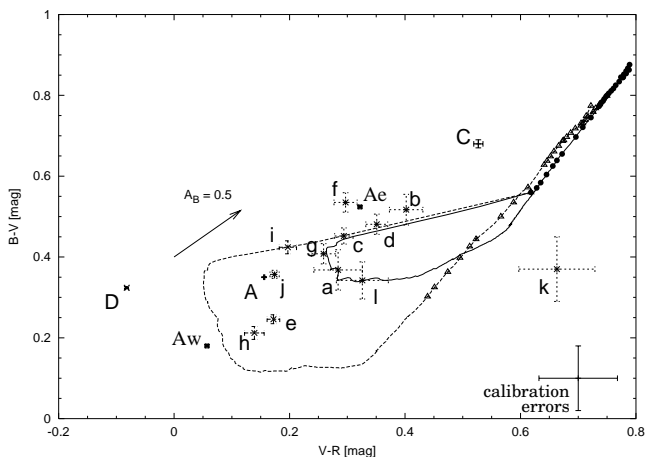}}
        \caption{Plot of the objects in field 1 of AM 0529-565. The
          models used are 4($Z_1$) (solid, $b=0.02$) and 6($Z_1$)
          (dashed, $b=0.18$). }
        \label{AM0529-565_plot}
\end{figure}

The image of this system at a redshift of $V_\mathrm{A} = 4420$ km
s$^{-1}$ is shown in Fig.~\ref{AM0529-565}.  The principal galaxy,
``A'', has an integrated absolute blue magnitude typical of a
Magellanic Irregular galaxy. However the presence of a double nucleus
and of diffuse tidal tails are characteristics of mergers.  The
eastern nucleus ``Ae'' has redder colors than the western one ``Aw''
which is currently in a starburst phase (see
Fig.~\ref{AM0529-565_plot}).  ``B'' is an elliptical galaxy. Given its
colors, it is probably a background object with an estimated redshift
of $\approx 0.2$ (M\"oller et al.~in prep.).  ``C'' is a peculiar low
surface brightness (LSB) galaxy of unknown redshift.  ``D'' is a blue
dwarf galaxy associated with ``A'' at a relative velocity of
$V_\mathrm{rel} = 220$ km s$^{-1}$.  Several diffuse knots can be seen
around ``A''; they are designated with ``a'' to ``l''. ``a'' seems to
be detached from the eastern tidal tail.
 
From preliminary spectroscopic test observations of AM 0529-565
(Weilbacher et al.~2000, in prep.), we could estimate an oxygen
abundance of $12+\log(O/H) = 8.1$ (equivalent to $Z = 0.003$) for
``e'' and ``h'' and a very low $12+\log(O/H) = 7.6$ (equiv.~to $Z =
0.0008$) for object ``D''. Because of its much smaller metallicity,
``D'' is most probably a pre-existing galaxy currently in interaction
with ``A''.
  
Fig.~\ref{AM0529-565_plot} shows that the colors of most of the knots
in the tidal debris are consistent with TDG burst models with low
metallicity ($Z_1$) and burst strength in the range $0.02 < b < 0.18$.

\subsection{\object{AM 0537-292}}
\begin{figure}
        \resizebox{\hsize}{!}{\includegraphics{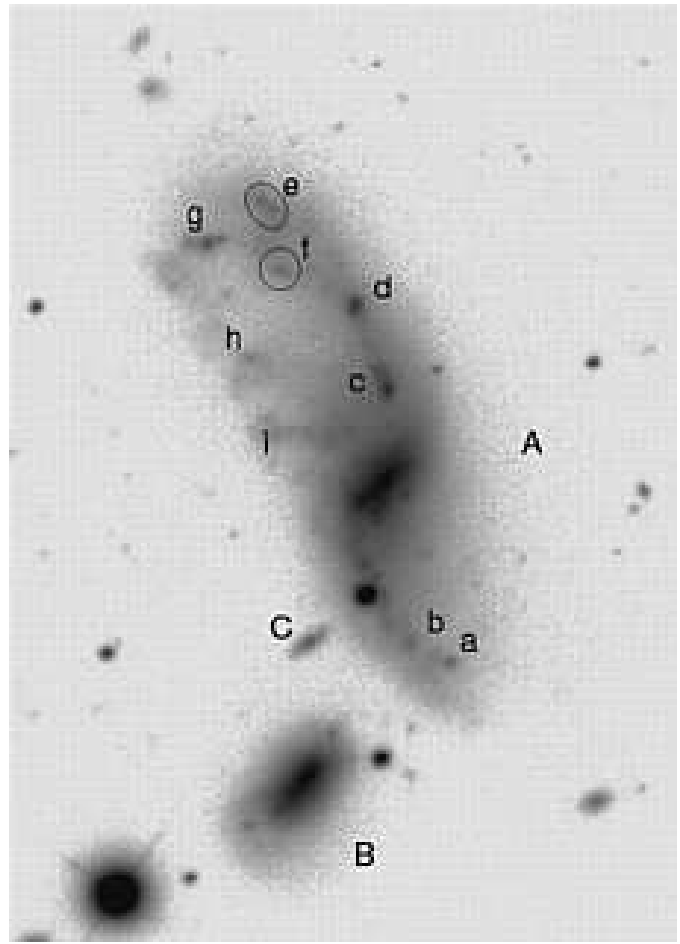}}
        \caption{Identification chart of field 2 around AM 0537-292.}
        \label{AM0537-292}
\end{figure}
\psfull 
\begin{figure}
        \resizebox{\hsize}{!}{\includegraphics{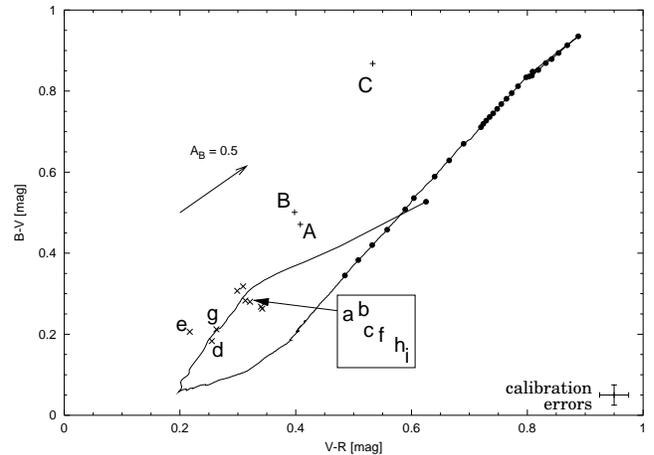}}
        \caption{Plot of the objects in field 2 of AM 0537-292. The
          model used is 5($Z_3$) (solid) with $b=0.09$. }
        \label{AM0537-292_plot}
\end{figure}

The image of this system also known as MCG-5-14-9 is shown in
Fig.~\ref{AM0537-292}.  ``A'' is a strongly disturbed galaxy with
tidal tails and a central bar.  ``B'' has the morphology of an
E5 galaxy in the central part, but the contours get distorted further
out. The overall color of both galaxies is strongly dominated by the
light of the nuclei.  ``C'' and all other unnamed visible extended
objects in the field are found to be background galaxies from their
colors.

Several large blue knots probably associated with star-forming regions
can be seen in the tails.  They are labeled ``a'' to ``i''.
Fig.~\ref{AM0537-292_plot} shows that all knots are well described by
the $Z_3$ model with a burst strength of the order of 10\%.  They have
similar burst ages of around 3 Myrs.

\subsection{\object{AM 0547-244}}
\begin{figure}
        \resizebox{\hsize}{!}{\includegraphics{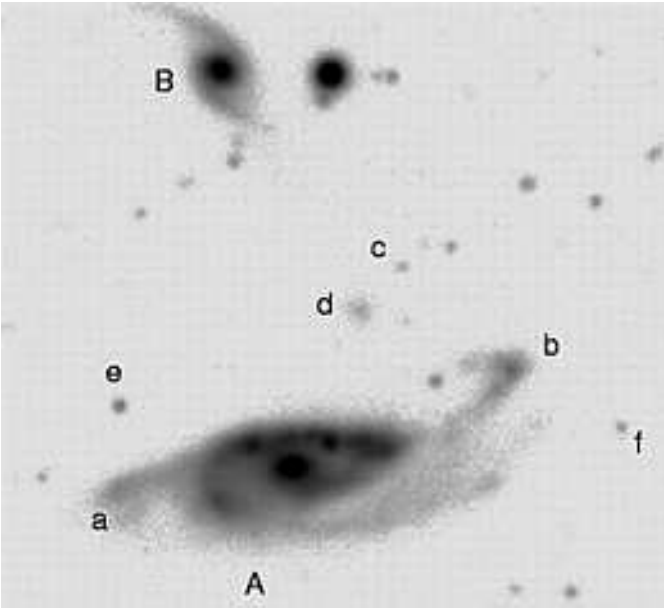}}
        \caption{Identification chart of field 3 around AM 0547-244.}
        \label{AM0547-244}
\end{figure}
\psfull 
\begin{figure}
        \resizebox{\hsize}{!}{\includegraphics{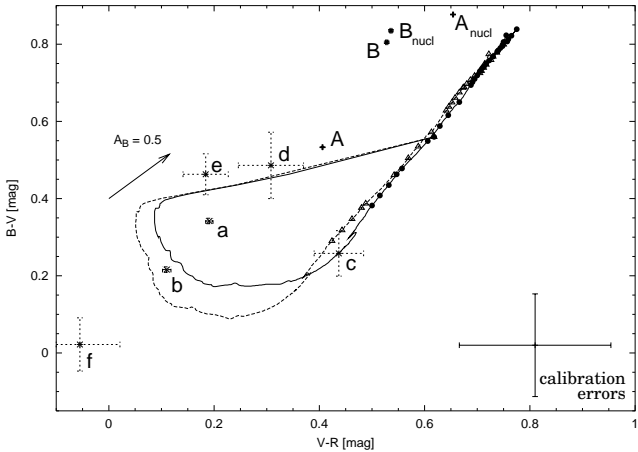}}
        \caption{Plot of the objects in field 3 of AM 0547-244. The
          models used are 5($Z_1$) (solid, $b=0.09$) and 6($Z_1$)
          (dotted, $b=0.18$).}
        \label{AM0547-244_plot}
\end{figure}

This system (Fig.~\ref{AM0547-244}) consists of a peculiar galaxy
``A'' ($V_\mathrm{A} = 13164$~km s$^{-1}$), from which three tidal
tails emanate, and a companion ``B'', the redshift of which is
unknown.  The colors of the nuclei of ``A'' and ``B'' are very red
(see Fig.~\ref{AM0547-244_plot}). This could indicate an `old' burst
age or strong reddening due to large amounts of dust.

At the ends of the tails of galaxy ``A'' two large condensations are
visible, denoted by ``a'' and ``b''. With absolute magnitudes of
$M_{B_\mathrm{a}} = -16.94$ and $M_{B_\mathrm{b}} = -17.05$, ``a'' and
``b'' are good candidates for luminous TDGs.  Around ``A'' four more
knots with unknown redshift can be seen (``c'' to ``f''). All other
objects have very red colors and are therefore most probably
background objects.

Fig.~\ref{AM0547-244_plot} shows that the objects ``a'' to ``d'' agree
well with the models with metallicity $Z_1$ and intermediate to high
burst strengths while ``f'' is not matched by any model. The
interpretation of ``a'' to ``f'' is severely limited by the large
systematic errors of this field.  A $Z_3$ metallicity model can
therefore not be excluded.

\subsection{\object{AM 0607-444}}
\begin{figure}
        \resizebox{\hsize}{!}{\includegraphics{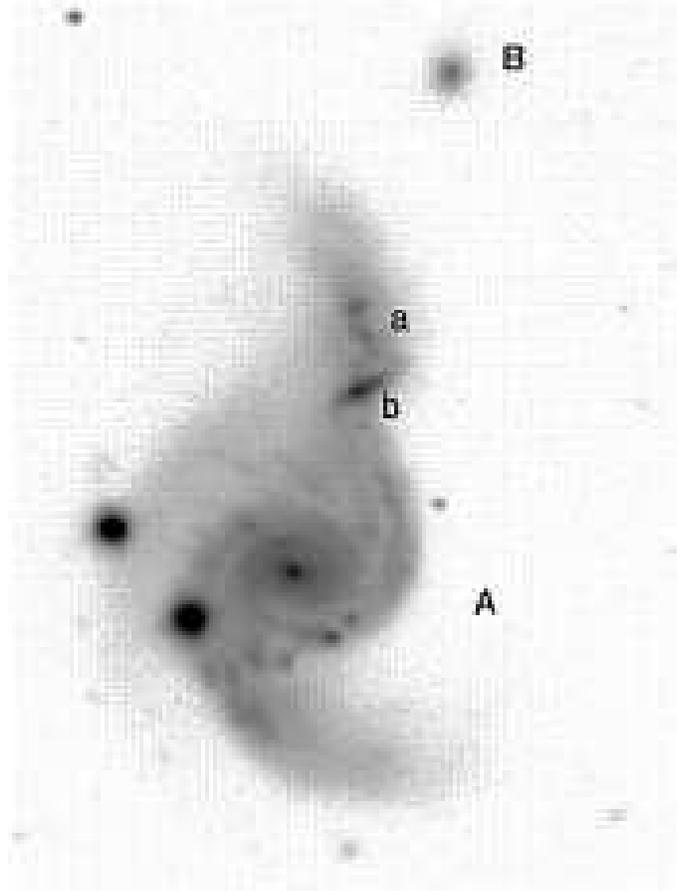}}
        \caption{Identification chart of field 4 around AM 0607-444.}
        \label{AM0607-444}
\end{figure}
\psfull
\begin{figure}
        \resizebox{\hsize}{!}{\includegraphics{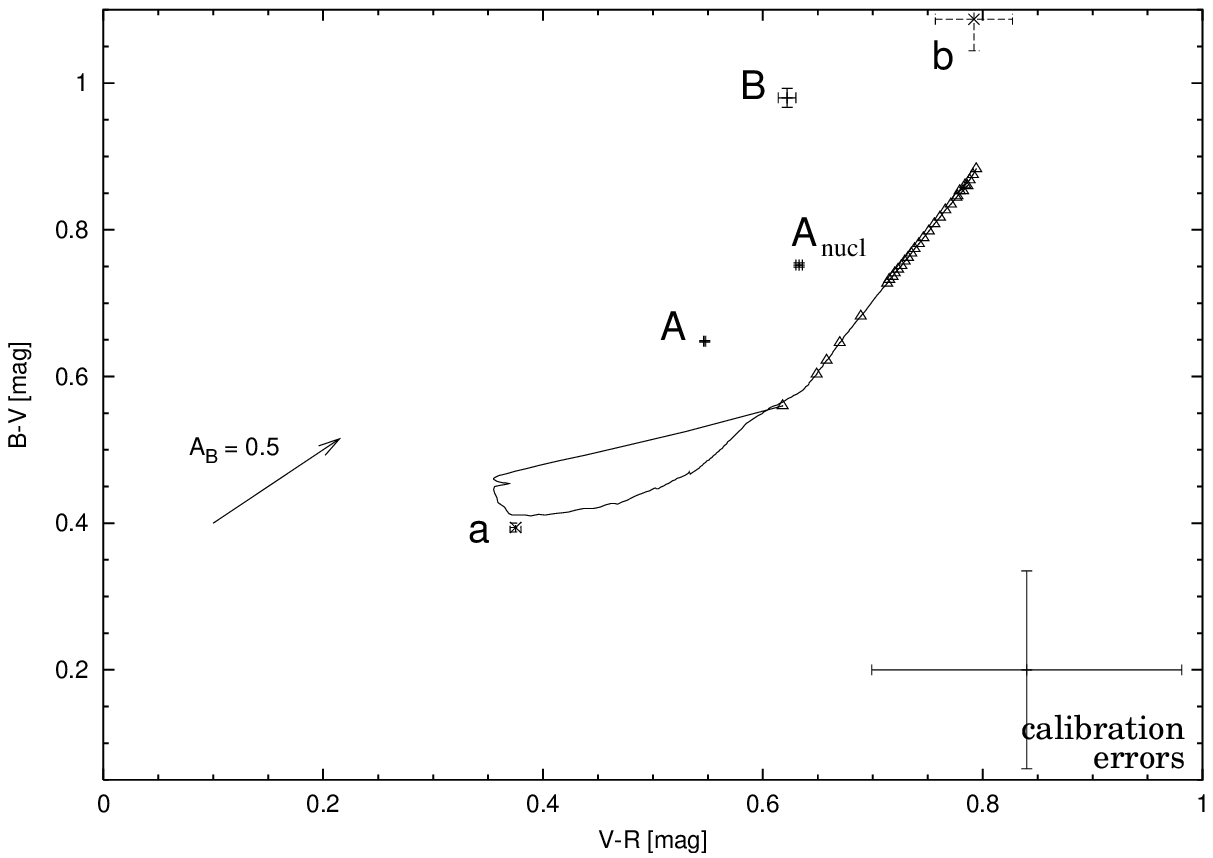}}
        \caption{Plot of the objects in field 4 of AM 0607-444. The
          model used is 1($Z_1$) (solid) $b=0.01$.  }
        \label{AM0607-444_plot}
\end{figure}

This system consists of a spiral galaxy (``A'') which is severely
disturbed in its outer parts (Fig.~\ref{AM0607-444}). The object
labeled ``B'' has red colors and may be a background galaxy.  Two
objects are seen towards the northern tail. The very blue knot at its
tip, ``a'', hosts a star-forming region. In contrast, ``b'' is much
redder, has elliptical contours in its center and may either be the
remnant core of the interacting companion of ``A'' or a background
galaxy superimposed on the tail.  The colors of the TDG candidate
``a'' are best matched with a $Z_1$ model with a weak burst
(Fig.~\ref{AM0607-444_plot}), note, however, the large systematic
errors of this field.

\subsection{\object{AM 0642-645}}
\begin{figure}
        \resizebox{\hsize}{!}{\includegraphics{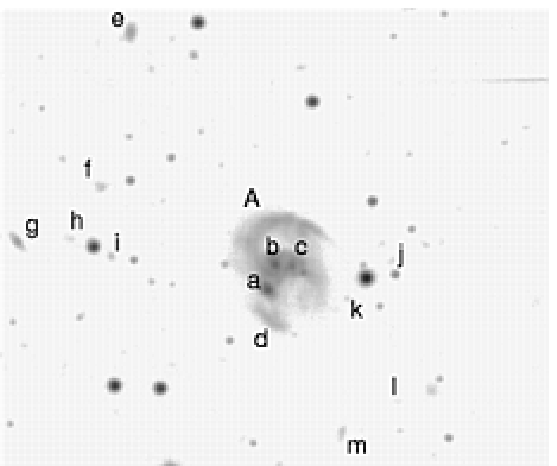}}
        \caption{Identification chart of field 5 around AM 0642-645.}
        \label{AM0642-645}
\end{figure}
\psfull 
\begin{figure}
        \resizebox{\hsize}{!}{\includegraphics{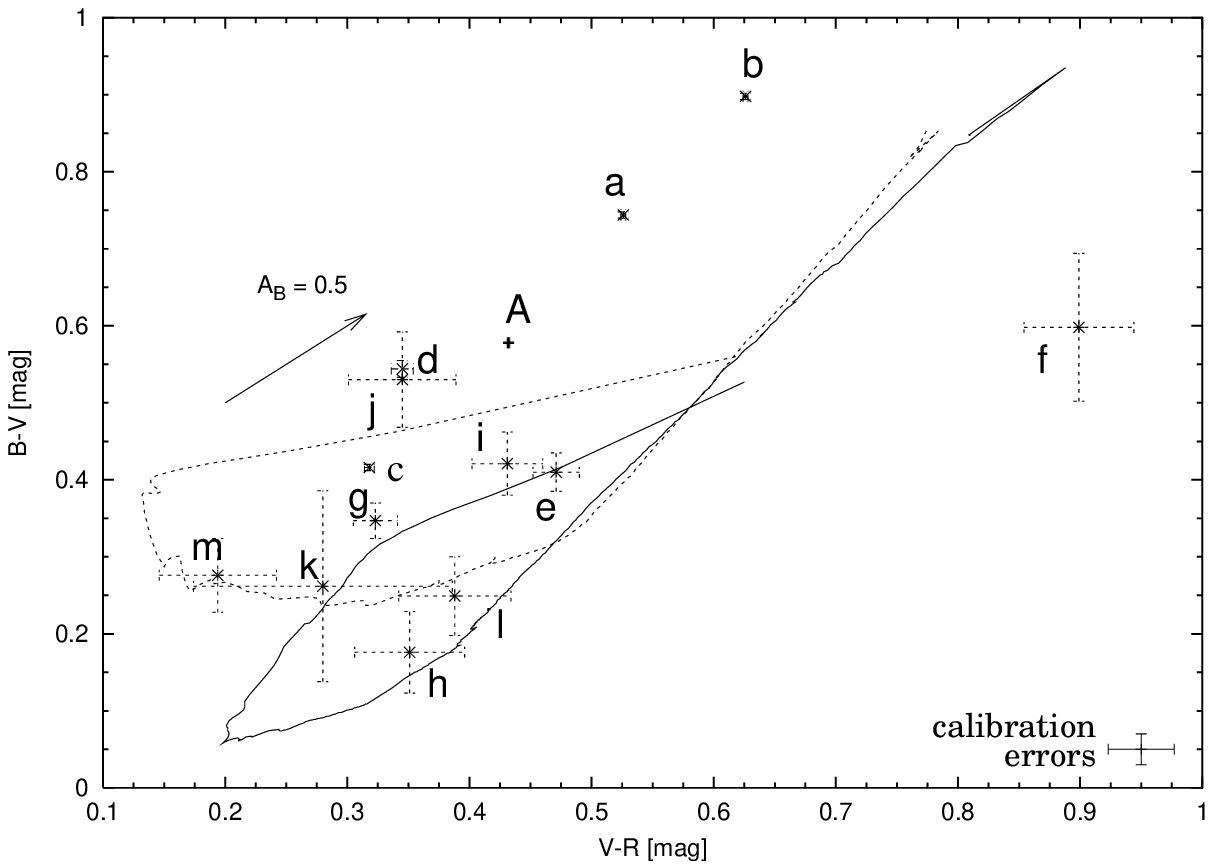}}
        \caption{Plot of the objects in field 5 of AM 0642-645. The
          models used are 2($Z_1$) (dotted, $b=0.05$) and 5($Z_3$)
          (solid, $b=0.09$).}
        \label{AM0642-645_plot}
\end{figure}

The peculiar system ``A'' (Fig..~\ref{AM0642-645}) exhibits two red
nuclei (``a'',``b'') and a bluer central knot,``c'', and may well be a
merger.  A plume to the southeast is designated with ``d'' and nine
other extended objects in the field of view that have colors in the
range of our models have been labeled.  Fig.~\ref{AM0642-645_plot}
shows that all of these objects would be consistent with TDG models
having weak to intermediate burst strengths with the exception of
object ``f'', which has colors similar to an elliptical galaxy at $z
\approx 1.3$ (M\"oller et al.~in prep.).  However, without a spectroscopic
redshift and in the absence of tidal structures linking them with the
parent galaxy, the tidal origin of the objects surrounding ``A'' is
very speculative.

\subsection{\object{AM 0748-665}}
\begin{figure}
        \resizebox{\hsize}{!}{\includegraphics{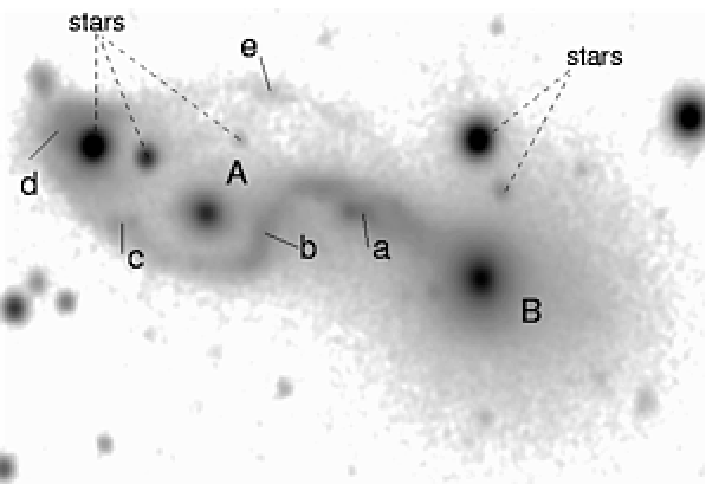}}
        \caption{Identification chart of field 6 around AM 0748-665.}
        \label{AM0748-665}
\end{figure}
\psfull
\begin{figure}
        \resizebox{\hsize}{!}{\includegraphics{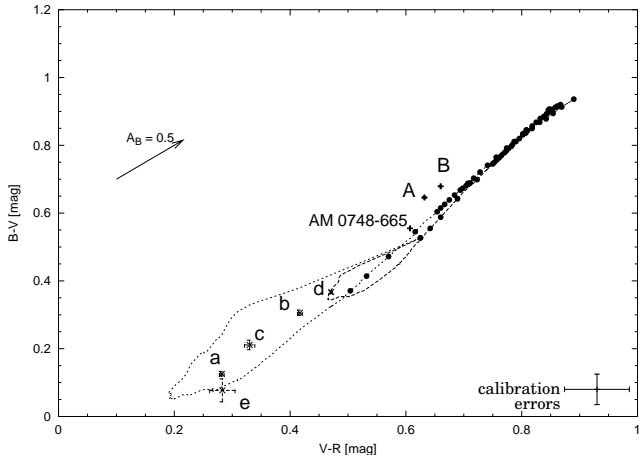}}
        \caption{Plot of the objects in field 6 of AM 0748-665. The
          models used are 1($Z_3$) (dashed) and 3($Z_3$) (dotted) with
          $b=0.01, 0.10$.}
        \label{AM0748-665_plot}
\end{figure}

This interacting system is composed of two galaxies (``A'' and ``B'')
which are connected by a long narrow bridge (Fig.~\ref{AM0748-665}).
Their morphology indicates that the progenitors of ``A'' and ``B'' may
have been a spiral and an elliptical, respectively. Several
condensations labeled ``a'' to ``e'' can be seen in the associated
tidal features.  The brightest condensation ``d'' is blended with a
bright star. The latter has been subtracted with a scaled PSF before
carrying out the photometry.  Fig.~\ref{AM0748-665_plot} shows that
all data points agree with $Z_3$ models that provide burst strengths
up to 0.10.

\subsection{\object{AM 1054-325}}
\begin{figure}
        \resizebox{\hsize}{!}{\includegraphics{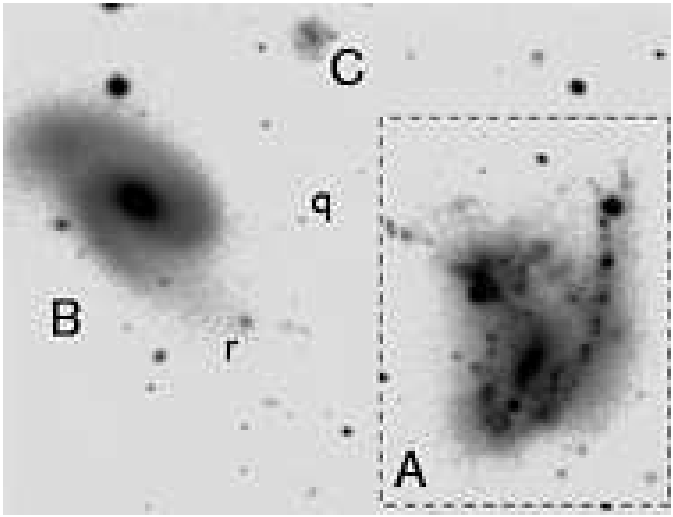}}
        \caption{Identification chart of field 7 around AM 1054-325.}
        \label{AM1054-325}
\end{figure}
\begin{figure}
        \resizebox{\hsize}{!}{\includegraphics{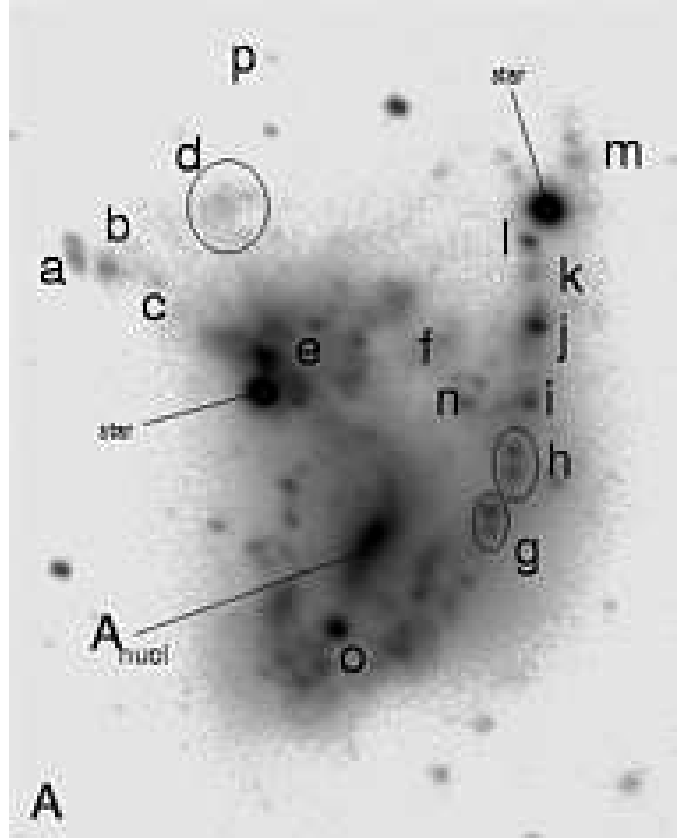}}
        \caption{Identification chart of AM 1054-325A.}
        \label{AM1054-325A}
\end{figure}
\psfull 
\begin{figure}
        \resizebox{\hsize}{!}{\includegraphics{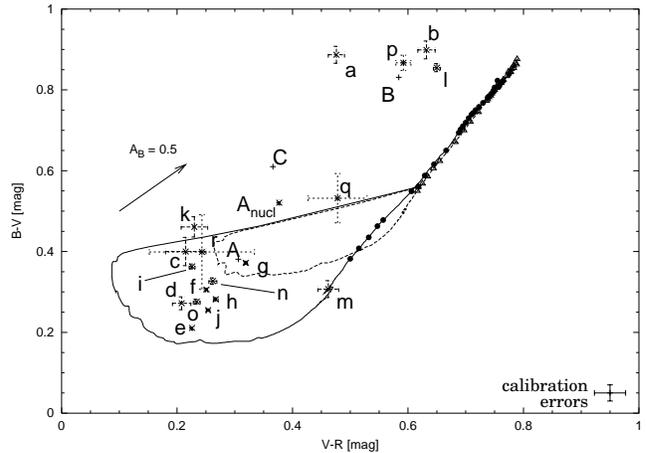}}
        \caption{Plot of the objects in field 7 of AM 1054-325. The
          models used are 4($Z_1$) (dashed, $b=0.02$) and 5($Z_1$)
          (solid, $b=0.09$).}
        \label{AM1054-325_plot}
\end{figure}

This presumably interacting system (Fig.~\ref{AM1054-325}) is composed
of an irregular \ion{H}{ii}-galaxy, ``A'', at a redshift of
$V_\mathrm{A} = 3795$ km s$^{-1}$ \citep{SW93} and a spiral-like
object ``B'', which has however colors that are redder and more
homogeneous than typical spiral galaxies. Its redshift is not known.
Object ``C'' is a LSB galaxy at an unknown redshift.

The enlarged image of ``A'' (Fig.~\ref{AM1054-325A}) indicates that
this galaxy seems to have two nuclei, one marked as
``A$_\mathrm{nucl}$'', the other as ``e'' (blended by a foreground
star). Like in AM 0529-565 A the two nuclei have different colors. The
former is redder ($B-V\!=0.52$) while the latter has the blue colors
of a strong starburst ($B-V\!=0.21$). \citet{PRM91} derived for ``A''
an oxygen abundance of $12+\log(O/H) = 8.14$ equivalent to $Z =
0.003$. We therefore use our $Z_1$ model for this galaxy.

In the western tail of ``A'' several large star-forming regions are
visible (``g'' to ``m'') with luminosities of fainter dwarf galaxies.
A few other features of special interest are marked as ``a'' to ``d''
and ``f''.  Fig.~\ref{AM1054-325_plot} shows that their colors are
well matched with $Z_1$ models with $0.02 < b < 0.09$.  However ``a'',
``b'' and ``p'' that have discrepant reddish colors are likely
background objects. Note that the outlier ``l'' has a very uncertain
photometry due to the vicinity of a bright saturated star.

\subsection{\object{AM 1208-273}}
\begin{figure}
        \resizebox{\hsize}{!}{\includegraphics{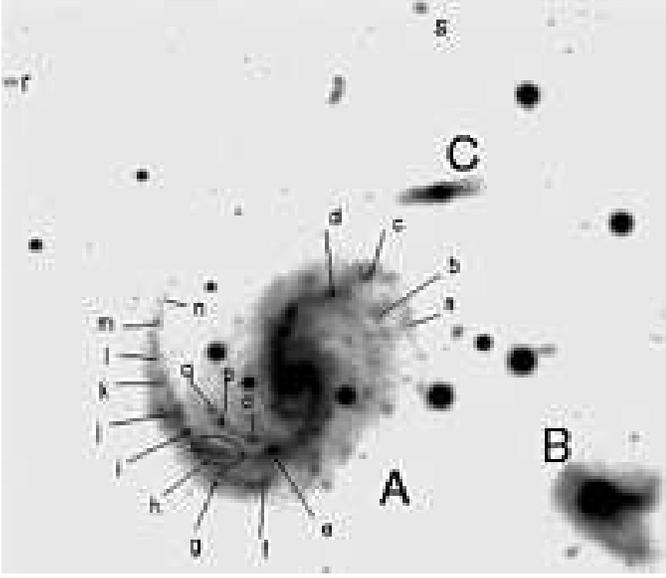}}
        \caption{Identification chart of field 8 around AM 1208-273.}
        \label{AM1208-273}
\end{figure}
\psfull 
\begin{figure}
        \resizebox{\hsize}{!}{\includegraphics{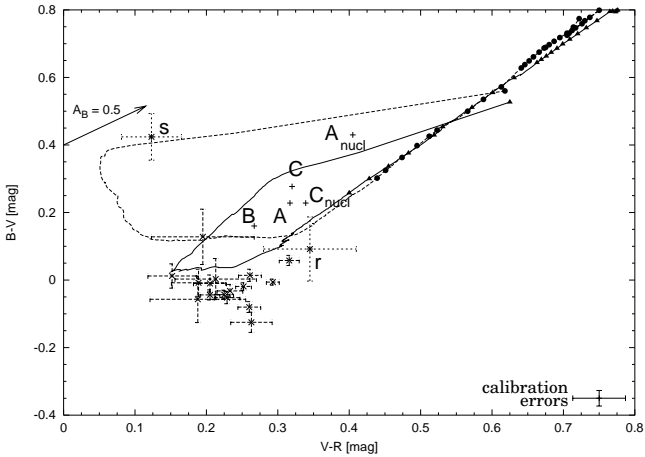}}
        \caption{Plot of the objects in field 8 of AM 1208-273. The
          models used are 6($Z_1$) (dashed) and 6($Z_3$) (solid) both
          with $b=0.18$.}
        \label{AM1208-273_plot}
\end{figure}

This system, also known as MCG-5-29-28, has three apparent members
(Fig.~\ref{AM1208-273}).  The main galaxy, ``A'', is a disturbed Sc
spiral with a small bar visible at the center ($V_\mathrm{A} = 14780$
km s$^{-1}$). ``B'' is a foreground object with an active starburst
\citep[$V_\mathrm{B} = 12433$ km s$^{-1}$,][]{DP97}. The redshift of
the disk galaxy ``C'' is not known.

Along the extended spiral arms of ``A'' several knots are visible,
marked as ``a'' to ``q''. Two detached diffuse objects ``r'' and ``s''
are visible northeast and north of the main galaxy, but their
association with the other galaxies is uncertain.
Fig.~\ref{AM1208-273_plot} shows that all knots in the arms are
matched by a $Z_3$ model with a high burst strength. They have roughly
the same burst age.

\subsection{\object{AM 1325-292}}
\begin{figure}
        \resizebox{\hsize}{!}{\includegraphics{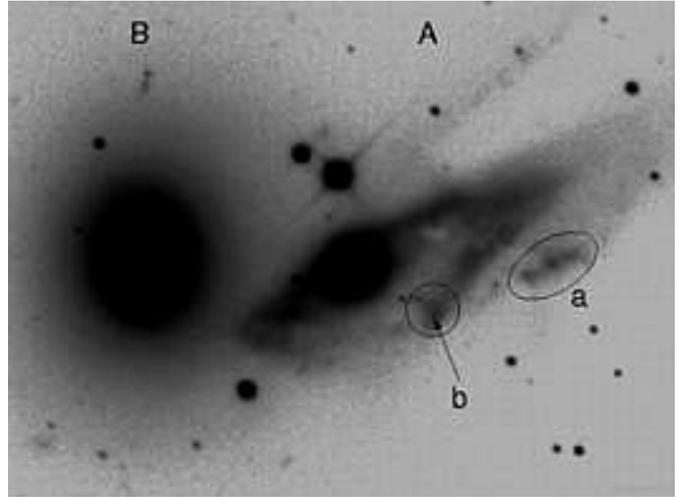}}
        \caption{Identification chart of field 9 around AM 1325-292.}
        \label{AM1325-292}
\end{figure}
\psfull
\begin{figure}
        \resizebox{\hsize}{!}{\includegraphics{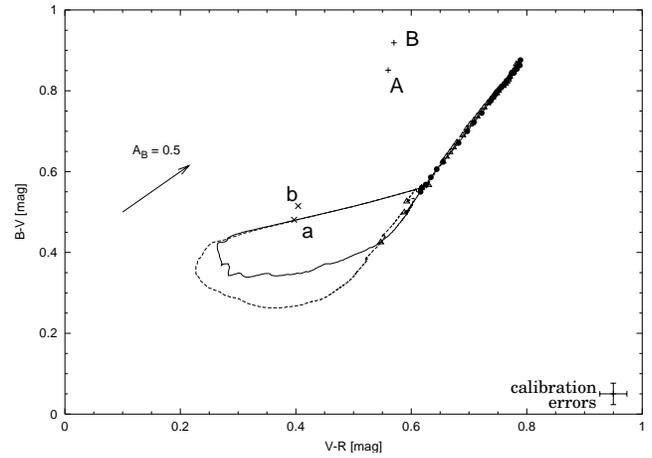}}
        \caption{Plot of the objects in field 9 of AM 1325-292. The
          models used are 4($Z_1$) (solid, $b=0.02$) and 7($Z_1$)
          (dashed, $b=0.04$).}
        \label{AM1325-292_plot}
\end{figure}

AM 1325-292 (Fig.~\ref{AM1325-292}) is a close interacting system
between a spiral, NGC 5152 (``A'') and an elliptical, NGC 5153
(``B''). Their redshifts are $V_\mathrm{A} = 4598$ km s$^{-1}$ and
$V_\mathrm{B} = 4431$ km s$^{-1}$, respectively.

The spiral has three visible spiral arms/tidal tails, a long one to
the north and two apparently shorter ones to the southwest, at the end
of which two bigger knots of dwarf galaxy size are seen (``a'' and
``b''). These knots have the bluest colors of the entire system and do
well agree with $Z_1$ models and small or intermediate burst strength
shown in Fig.~\ref{AM1325-292_plot}. The small photometric errors
preclude $Z_3$ models although due to the nature of the parent galaxy
a higher metallicity would have been expected.

\subsection{\object{AM 1353-272}}
\begin{figure}
        \resizebox{\hsize}{!}{\includegraphics{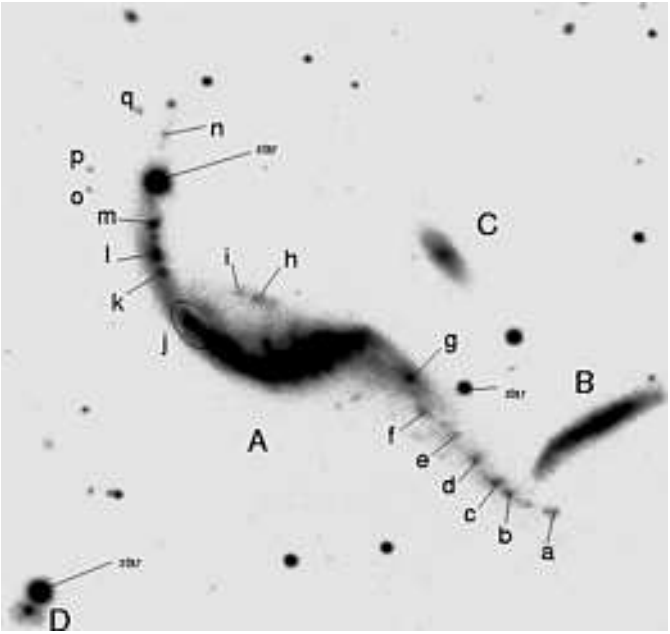}}
        \caption{Identification chart of field 10 around AM 1353-272.}
        \label{AM1353-272}
\end{figure}
\psfull 

Four galaxies are seen in the field of AM 1353-272
(Fig.~\ref{AM1353-272}).  The two brightest are at the same redshift.
``A'' is an integral-sign spiral ($V_\mathrm{A} = 12145$ km s$^{-1}$)
interacting with a disturbed disk galaxy ``B'' ($V_\mathrm{B} = 11791$
km s$^{-1}$). ``C'' is most probably a background elliptical, as
judged from its red colors. ``D'', to the southeast, is a LSB galaxy.
Its color is highly uncertain due to the contamination by a bright
foreground star.

The long tidal tails of ``A'' host a number of blue knots (``a'' to
``n'').  Three detached small knots (``o'' to ``q'') are visible east
of the northeast tail.  Given their colors, these knots could be
physically linked with ``A'' but without redshift information
available this remains speculative.  The knots have absolute
magnitudes in the range of $-12.1 < M_B < -15.7$ mag. Their colors do
agree with the $Z_3$ models and intermediate strength burst.
Condensations ``d'' and ``m'' have however deviant colors that seem
more consistent with $Z_1$ models.

\section{Discussion}\label{Disc-Sec}
\subsection{Background contamination}\label{BgCont-Sect}
In each field a number of faint extended objects are visible (see
e.g.~Figs.~\ref{AM0537-292} and \ref{AM0547-244}) that have red
colors. They lie outside the color range of our models that are valid
for $z = 0$ only. Most of them should hence be background objects.
This was tested using the chemically consistent evolutionary synthesis
models of \citet{MFF98} and M\"oller et al.~(in prep.), which include
cosmological parameters and give colors of all galaxy types as a
function of $z$. A rough photometric redshift could be determined from
the $B$,$V\!$,$R$ measurements.  The few faint objects with colors
that strongly deviate from our TDG models and that are inconsistent
with those of distant galaxies could be red dwarf galaxies
pre-existing the collisions. They have been excluded from our
analysis.  One SUSI field with an area of $1.3\cdot10^{-3}\ \sq\degr$
contains on average 7.0$\pm$0.7 and 13.0$\pm$4.0 objects with deviant
colors in the respective magnitude ranges $B = 22\dots 23$ and $B =
23\dots 24$.  After correction for the occlusion of background objects
due to the interacting system (typically $2\cdot10^{-4}\ \sq\degr$),
these values are compatible with those of \citet{MSF91}.  They imaged
12 fields with an area of $6.8\cdot10^{-3}\ \sq\degr$ and counted
$5000\pm800$ and $15000\pm1000$ galaxies, respectively, for the
magnitude ranges in question. For our field of view, the numbers would
be 6.8$\pm$1.1 and 20.2$\pm$1.3.  Therefore we do not see an
overdensity of background galaxies in our fields contrary to what
\citet{DMT98} found in their sample of interacting systems. Their
field of view is much larger (about $3.3\cdot10^{-3}\ \sq\degr$) than
ours and they have better statistics.  But they mostly imaged
interacting systems with lower $z$ which hence occlude a larger
portion of the image. Neither did they use color information to
preselect possible nearby objects. Moreover, they mainly found an
overdensity in the brightness range $R=18\dots 19.5$ mag, where we do
not have enough background objects for statistical analysis.

\subsection{Properties of the TDG candidates}\label{TDG-Sec}
As seen before, several knots turn out to be background objects as
judged from their colors. One should hence be cautious when physically
linking faint objects with apparently nearby large galaxies. This is
especially true for the detached objects lying far away from the tidal
tails of the parent galaxies. In the absence of a redshift
measurement, the color information and its comparison with the model
predictions is useful to disentangle between possible objects in or
close to tidal features from background galaxies.  Using this
technique, we have compiled a catalog of candidates for TDGs or their
progenitors, i.e.~knots associated with or close to tidal features
that -- within the observational errors -- agree with one or more of
our evolutionary synthesis models.  They are listed in Table
\ref{AllTDGCand-Tab}.

Column 1 shows the designation of the candidate.  Column 2 gives, when
its distance is known, the absolute blue magnitude of the TDG
candidate. Columns 3 and 4 resp.~give its $B-V\!$ color and the
difference in $B-V\!$ between the knot and the surrounding tail
(values in parentheses indicate detached knots). The last three
columns list the results from the comparison with the models.  Column
5 gives the ID of the best matching model(s), column 6 the most
probable burst age(s), and column 7 a range of possible burst
strengths.  Note that the best-fit models shown here for every
individual TDG candidate are not necessarily those shown in the plots
in Sect.~\ref{Results-Sec}, where models were chosen to represent a
maximum number of data points.

\begin{table*}[tbp]
\caption{TDG candidates}
\label{AllTDGCand-Tab}
    \begin{tabular}{l | c c c | c c c}
    \textrm{Designation}&$M_B$&$B$\,$-$\,$V$&\hspace*{-5pt}$\Delta (B-V)$&Best fit &\hspace*{-5pt}burst age & burst strength\\
                  &  [mag]& [mag] &       \hspace*{-5pt}(tail-TDG)  & Model ID & [Myr]                    &   $b$ \\
    \hline
    AM 0529-565a & -11.48 & 0.37  &  (0.03)&                          4($Z_1$) &          4 & 0.02 \\
    AM 0529-565b & -10.65 & 0.52  & (-0.12)&                          6($Z_1$) &       $<$1 & 0.01..0.18 \\
    AM 0529-565g & -11.64 & 0.41  &   0.01 &                          4($Z_1$) &          3 & 0.02..0.05 \\
    AM 0529-565i & -12.57 & 0.42  &   0.15 &                          6($Z_1$) &       $<$1 & 0.05..0.18 \\
    \hline
    AM 0537-292a &        & 0.31  &   0.15 &                          6($Z_3$) &          1 & 0.04..0.18 \\
    AM 0537-292d &        & 0.18  &   0.28 &                          5($Z_3$) &     10$/$3 & 0.05..0.18 \\
    AM 0537-292e &        & 0.21  &   0.20 &                          5($Z_1$) &      8$/$3 & 0.09..0.18 \\
    AM 0537-292f &        & 0.28  &   0.11 &                          5($Z_3$) &$\approx 1$ & 0.05..0.10 \\
    AM 0537-292g &        & 0.21  &   0.16 &                          3($Z_3$) &     10$/$3 & 0.09..0.18 \\
    \hline
    AM 0547-244a & -16.94 & 0.34  &   0.07 &                          9($Z_1$) &          4 & 0.04..0.05 \\
    AM 0547-244b & -17.05 & 0.22  &   0.12&                           3($Z_1$) &    $<$1..4 & 0.09..0.18 \\
    AM 0547-244d & -14.46 & 0.49  & (-0.15)&                          6($Z_1$) &       $<$1 & 0.02..0.18 \\
    \hline
    AM 0607-444a &        & 0.39  &   0.34 &                          1($Z_1$) &   $<$1$/$6 & 0.01..0.18 \\
    \hline
    AM 0748-665a &        & 0.13  &   0.32 &                          2($Z_3$) &     3$/$10 & 0.05 \\
    AM 0748-665d &        & 0.21  &0.11$^1$&                          1($Z_3$) &         20 & 0.01 \\
    AM 0748-665e &        & 0.08  &   0.15 &                          3($Z_3$) &         10 & 0.05..0.18 \\
    \hline
    AM 1054-325g & -14.29 & 0.37  &   0.08 &                          4($Z_1$) &          5 & 0.02 \\
    AM 1054-325h & -13.73 & 0.28  &   0.11 &                          7($Z_1$) &          6 & 0.04 \\
    AM 1054-325i & -13.25 & 0.36  &  -0.02 &                          7($Z_1$) &          4 & 0.04 \\
    AM 1054-325j & -14.36 & 0.26  &   0.12 &                          2($Z_1$) &          6 & 0.05 \\
    AM 1054-325k & -10.40 & 0.46  &  -0.14 &                          6($Z_1$) &       $<$1 & 0.05..0.18 \\
    AM 1054-325m & -11.40 & 0.31  &0.08$^1$&                          5($Z_1$) &         70 & 0.05..0.18 \\
    \hline
    AM 1325-292a & -15.30 & 0.48  &   0.15 &                          5($Z_1$) &       $<$1 & 0.01..0.18 \\
    AM 1325-292b & -15.46 & 0.52  &   0.18 &                          6($Z_1$) &       $<$1 & 0.01..0.18 \\
    \hline
    AM 1353-272a & -13.68 & 0.13  & (0.19) &                          7($Z_3$) &         40 & 0.04..0.10 \\
    AM 1353-272b & -13.10 & 0.24  &   0.07 &                          2($Z_3$) &      9$/$3 & 0.04..0.18 \\
    AM 1353-272c & -13.58 & 0.24  &   0.02 &                          2($Z_3$) &          8 & 0.05 \\
    AM 1353-272d & -13.98 & 0.29  &   0.10 &                          2($Z_1$) &          4 & 0.05 \\
    AM 1353-272e & -14.03 & 0.11  &   0.27 &                          5($Z_3$) &         30 & 0.05..0.10 \\
    AM 1353-272f & -12.84 & 0.33  &   0.05 &                          4($Z_3$) &  $<$1$/$20 & 0.02..0.18 \\
    AM 1353-272h & -14.01 & 0.34  &   0.21 &                          6($Z_3$) &     2$/$10 & 0.04..0.18 \\
    AM 1353-272i & -12.13 & 0.44  &   0.04 &                          1($Z_3$) &  $<$1$/$40 & 0.01..0.18 \\
    AM 1353-272k & -13.99 & 0.20  &   0.11 &               5($Z_1$)$/$6($Z_3$) &      8$/$3 & 0.05..0.18 \\
    AM 1353-272l & -15.25 & 0.17  &   0.12 &               5($Z_1$)$/$6($Z_3$) &      9$/$3 & 0.09..0.18 \\
    AM 1353-272m & -14.49 & 0.14  &   0.03 &                          2($Z_1$) &          4 & 0.05..0.10 \\
    AM 1353-272n & -14.31 & 0.17  &   0.13 &                          2($Z_3$) &     10$/$3 & 0.05..0.10 \\
    \hline
    \end{tabular}\\
$^1$Contamination by a nearby bright star.
\end{table*}

\subsubsection{Number and luminosity}
\begin{figure}
  \resizebox{\hsize}{!}{\includegraphics{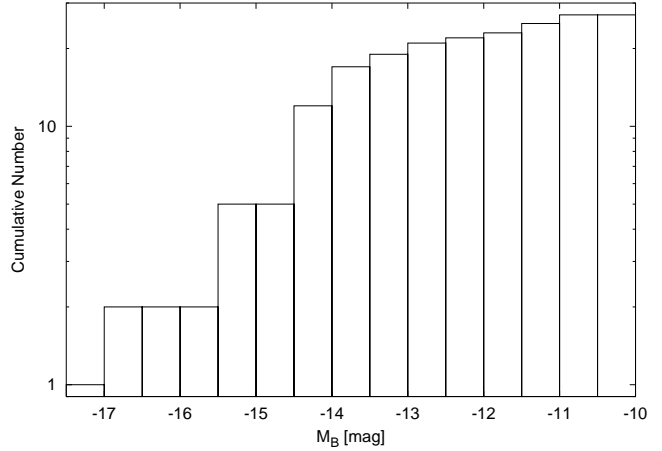}}
  \caption{Luminosity function of the tidal objects.}
  \label{LumFunc-Fig}
\end{figure}

In our sample of ten, 8 interacting systems contain promising TDG
candidates.  On average 3.6 were identified per system.  The
luminosities of the candidates for which the distance of their parent
galaxies is known range from faint ($M_B = -10.4$ mag) to brighter
dwarf galaxies ($M_B = -17.1$ mag) with a mean of $\left<M_B\right>=
-13.6$ mag.  It should be noted that even though we have some faint
TDG candidates in our sample, they are still more luminous than normal
\ion{H}{ii} regions in spiral galaxies by about 4 mag \citep{BK97}.
The luminosity function of the TDG candidates is shown in
Fig.~\ref{LumFunc-Fig}.

\subsubsection{Colors and star formation}
\begin{figure}
  \resizebox{\hsize}{!}{\includegraphics{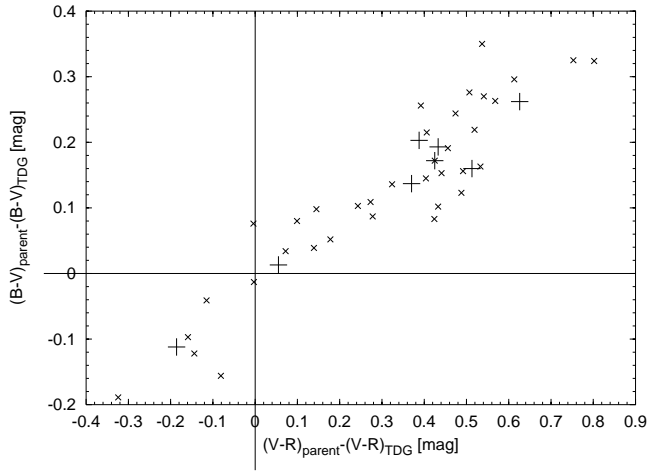}}
  \caption{Two-color-diagram of the differential colors of parent
    galaxies and TDG candidates. The small crosses mark the TDG
    candidates selected in Table~\ref{AllTDGCand-Tab}. The big crosses
    mark the mean value of all TDG candidates in one given system.}
  \label{AllDiff-Fig}
\end{figure}

In Fig.~\ref{AllDiff-Fig} we compare the colors of these knots with
the overall colors of their parent galaxy. Note that the measurement
of the parent galaxy includes the light of the knots, but this does
not have a strong effect as the knots only give a neglegible
contribution (up to 1\% of the flux) to the brightness of the galaxy.

On average the TDG candidates have bluer colors than their parent
galaxies, which could be due to a number of effects:
\begin{itemize}
\item Fading effect: only the bluest clumps are visible.  As we have
  selected the candidates from their surface brightness in the
  $V\!$-band images and all candidates have a young burst age, it is
  possible that after a longer burst the objects become redder and
  fade in luminosity and are no longer visible as surface brightness
  peaks in $V$.
\item High relative gas content: the knots may experience stronger
  starbursts than the parent galaxies. A prerequisite for this could
  be a high gas content in the tidal features as observed in other
  systems \citep[e.g.][]{DM98}.
\item Time difference: The burst in the parent galaxies may take place
  before the star formation in the tidal features starts. The color of
  the parent galaxy would then be dominated by the color of the red
  central part, especially after a strong nuclear starburst has faded.
\item Smaller dust content: If the knots have a smaller dust
  extinction than the parent galaxies, their colors would appear
  redder. 
\item The colors of the parent galaxies are largely dominated by the
  colors of their nuclei which are usually redder than the disk from
  which a TDG may form.
\end{itemize}

\begin{figure}
  \resizebox{\hsize}{!}{\includegraphics{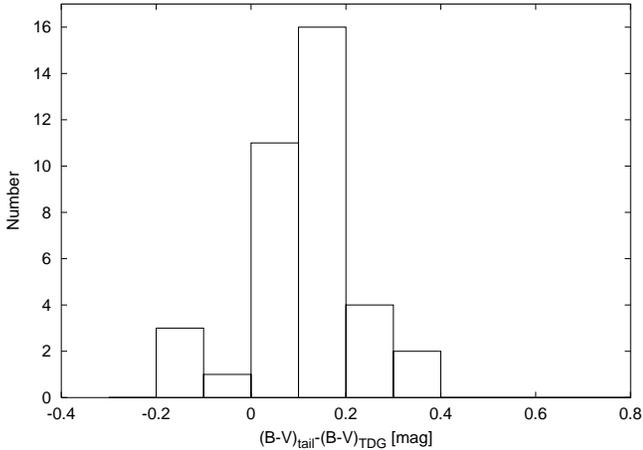}}
  \caption{Histograms of the differential colors of tidal features 
    and the TDG candidates in $B-V\!$.}
  \label{TailDiff-Fig}
\end{figure}

In order to tackle the last bias, we have compared the colors of the
knots with the color of their immediate surroundings, i.e.~in the
tails.  For knots already detached from the visible tail, the nearest
part of the tail was taken for the measurement.  The results are shown
in Table~\ref{AllTDGCand-Tab}, column~4 and in Fig.~\ref{TailDiff-Fig}.
Nearly all knots have a $B-V\!$ color bluer than their surrounding
tidal features.  We therefore conclude that the knots are actively
forming stars
while their surrounding material is more quiescient. \\
As already noticed by \citet{SWS90}, the tidal features and especially
the surface brightness peaks which we identify as TDG candidates often
show the bluest $B-V\!$ colors of the entire interacting system.

Some examples of interacting galaxies show star formation in the tails
but already have postburst characteristics in the central part of the
merger \citep[e.g.][]{HGvG94}.  One could therefore conclude that star
formation propagates from the central part to the outer regions along
the tails and look for such an effect in interacting systems with
several condensations along the tails. We do not find, however, any
clear trends in the properties of the knots along the tails in the two
systems which have longer tails (AM 1054-325A and AM 1353-272A)
neither in luminosity or color, nor in burst strength or burst age.

\subsubsection{Burst properties}
The model bursts developed in this study that best fit the TDG
candidates require intermediate to strong burst strength ($0.01 < b <
0.18$). Taking into account possible internal extinction would favor
even stronger bursts.  These burst strengths therefore seem to be
higher than for BCDGs, which have typical strengths of $b \approx
0.01$ or less \citep{Krue92}.  The burst strengths of our TDG
candidates also indicate that the mass is still governed by the old
stars ($b < 0.20$), but the luminosities in $B$ and $R$ are already
dominated by the young component (see Table~\ref{ModelGrid-Tab}).

The mean burst age of the tidal objects is about 8 Myrs.  Their
dynamical timescale may be determined assuming a typical distance of
the knots from the nucleus of the parent galaxy of 15 kpc -- about
half of the length of the tails of AM 1353-272 -- and a typical
rotation velocity in a spiral galaxy of 150 km s$^{-1}$.  With these
values, disk material is expelled to the present positions of the
knots in about 100 Myrs. Only very few objects in our sample show
comparable high burst ages. This suggests that the star formation
episode in the TDG candidates has started in situ in the tidally
expelled material.

\subsection{Nature of the TDG candidates}
The condensations in the tidal tails which we call ``Tidal Dwarf
Galaxy candidates'' show properties in the range expected for TDGs.
We call them {\it candidates}, because they still lack spectroscopic
confirmation and especially the proof that they are bound objects.
  
The TDG candidates identified in the systems presented here are on
average less luminous but more numerous than the TDGs found in other
interacting systems studied up to now. These other TDGs have an
absolute blue magnitude of -15.5 mag on average, part of them have
been found to be kinematically decoupled from their tails
\citep{DBW97,DM98}, and one or two of them are produced per collision.
Moreover, the TDGs known to be gravitationally bound are found at the
tip of the optical tails whereas our TDG candidates are distributed
all along the tidal features without any trend in luminosity in
agreement with one of the models discussed above.  One reason for
these differences may lie in the morphologies of the parent galaxies.
TDGs have so far essentially been observed in merging systems
involving massive spirals. Some of the interacting systems presented
here are composed of late type galaxies, some of which have
luminosities low enough to be classified as dwarf galaxies (e.g.~AM
0529-565).  It thus does not seem surprising that they form less
massive condensations.
  
\subsubsection{Formation of bound objects}
Numerical models of galaxy collisions show the presence of bound
objects in tidal debris that have luminosities and distributions quite
similar to the TDG candidates studied in our sample
\citep[e.g.][]{BH92}. The physical process for their formation is
still largely unknown. They might either form from growing
instabilities in the stellar component of tidal debris that may or may
not attract surrounding gas \citep{BH92}, or alternatively from
expelled gas that collapses and forms stars as proposed by
\citet{EKT93}.  The blue colors of our TDG candidates favor the second
scenario. Indeed we did not find any evidence for the presence for
pure stellar condensations in the tidal features of our interacting
systems. Our models show that recent star formation with a minimum
strength of $b \approx 0.01$, i.e.~forming 1\% of young stars is
necessary to account for the colors of our TDG candidates.  In most
cases, however, a much stronger burst is required.  Note however, that
we cannot distinguish between a stellar condensation into which gas
falls later to cause the starburst and an initially gaseous
condensation that forms stars on an underlying population of old
stars.
  
\subsubsection{Future evolution}\label{FutEvol-Sect}
\begin{figure}
  \resizebox{\hsize}{!}{\includegraphics{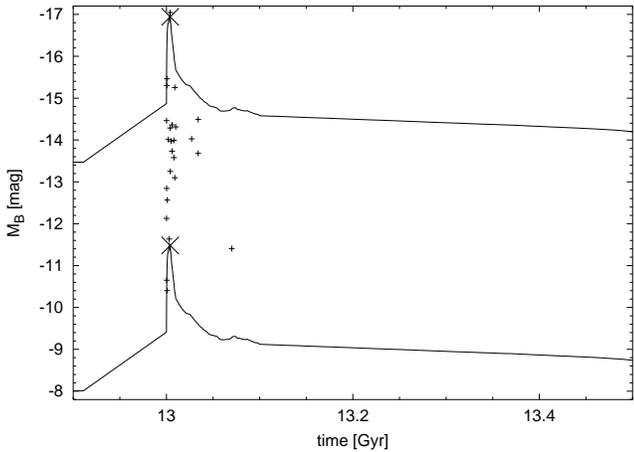}}
  \caption{Luminosity evolution of the selected TDG candidates which
    are displayed as crosses according to their luminosity and burst
    age.  The models shown are 06($Z_1$) with $b=0.18$ shifted in
    $M_B$ to fit the two candidates represented by the larger crosses.
    See text about discussion of these two examples.}
  \label{LumEvol-Fig}
\end{figure}

The spectrophotometric models are also used to extrapolate the
evolution of the knots, i.e.~to predict how luminosities and colors
will change with time. The latter is already visible on the two-color
diagrams in Sect.~\ref{Results-Sec}, where the evolution is marked in
steps of 100 Myr along the model curves.  Typically one Gyr after the
burst, the optical colors of the condensations will become
indistinguishable from the pre-burst colors unless further star
formation episodes take place.  The luminosity evolution is shown in
Fig.~\ref{LumEvol-Fig}, where models for one of the most luminous TDG
candidates, AM 0547-244a, and one of the least luminous candidates, AM
0529-565a, are displayed.  Without secondary bursts, the
$B$-luminosity will decrease by about 2.5 mag within 200 Myr from the
present age.  AM 0529-565a, with a present day absolute magnitude of
$M_B = -11.48$, will fade to $M_B\approx -8.0$ mag within 2 Gyrs.  It
will then have a luminosity similar to those of the faintest known
galaxies of the local group \citep{Mat98}. It would become invisible
in our images at the distance of AM 0529-565.  On a similar timescale,
the more luminous TDG candidate AM 0547-244a ($M_B = -16.94$) will
fade to $M_B \approx -13.4$ mag. It would be among the brighter dwarfs
in the local group and could still be observed at its actual distance
($m_B \approx 22.8$ mag).

The formation of objects as massive as TDGs in tidal debris is still
puzzling. One may wonder whether the numerous knots that we observe
here and that are produced in numerical simulations of galaxy
collisions could be the progenitors of the more massive TDGs
identified before.  As the photometric models indicate a rapid fading
of the TDGs after the burst, the evolution of a TDG progenitor into a
TDG would require that star formation in these condensations is
sustained for at least 100 Myrs, supposing that the progenitors
initially have a large enough gas reservoir. TDG progenitors could
grow that way and reach masses and luminosities of the well known
TDGs.  On the other hand, by that time, some of the less active TDG
progenitors might have faded and those initially situated close to
their parent galaxies might have fallen back, explaining why classical
TDGs are found preferentially at the end of tidal tails.  Observations
of a larger sample of systems and more detailed simulations would be
necessary to confirm the link between our TDG candidates and more
luminous TDGs.
  
Another possibility is that the TDG candidates in this paper do not
evolve into larger TDGs.  Instead some of our TDG candidates may not
grow much and eventually form more fragile TDGs of lower luminosity.
Even if fading after the burst is taken into account, our faintest TDG
candidates still have luminosities comparable to those of faint local
group dwarf galaxies.  The question if they will be able to survive as
bound and dynamically independent objects can only be answered by
observations of the kinematics in combination with detailed dynamical
models.

To be able to do more than speculate about the dynamical fate of the
TDG candidates, kinematical data of the system and the mass of the
gaseous component of each knot is required. To further restrict the
mass of the underlying old component NIR observations will be needed.

\section{Conclusions}\label{Conc-Sec}
We have presented imaging data of 10 southern interacting systems from
the Arp \& Madore Catalog selected for their resemblance to disturbed
distant galaxies. Our high quality images reveal a heterogeneous set
of morphologies from loosely interacting dwarf galaxies to disk-disk
mergers.  Through detailed photometry based on polygonal apertures in
the three optical bands $B$,$V\!$,$R$ we were able to measure the
colors of about one hundred condensations in or near the tidal
features of the interacting galaxies.

We have computed a grid of evolutionary synthesis models simulating
starbursts of various strengths in an underlying stellar population of
mixed ages typical of spiral galaxies. Such models reproduce well the
situation of TDGs and their progenitors that are composed of an old
population pulled out from the parent galaxy and young stars formed in
situ.  Our models take into account the emission lines and the
continuum from the photoionized gas with appropriate metallicities.
Comparing our photometric data with these models, we selected 36 blue
condensations as possible TDGs.  Follow-up spectroscopy is required in
some cases to assess their physical association with the interacting
system.

We discussed the overall properties of the 36 TDG candidates, noting
that they have luminosities brighter than \ion{H}{ii} regions in
spiral galaxies and more typical for that of dwarf galaxies.  Their
color is much bluer than those of their parent galaxies and mostly
bluer than the surrounding material in the tidal features. The
contribution of young stars to their mass is at least 1\% and may
reach 20\% or more for the most active condensations. In any case
their $B$-band luminosity is dominated by this young population. The
models also indicate that after the starburst when the young stars
cease to be the dominating source the condensations may fade by up to
2.5 mag in $B$ within 200 Myrs after the burst. Although the TDG
candidates seem to have young burst ages, we did not find any case of
condensations solely made of old stars from the parent galaxies
although the formation of such objects is also predicted in numerical
simulations.
  
Finally the TDG candidates studied here have properties different from
those of the more massive TDGs typically found at the tip of giant
tidal tails of nearby interacting systems.  We discuss whether our TDG
candidates could be progenitors of TDGs or a class of less luminous
objects.

\begin{acknowledgements}
  We wish to thank the team at the NTT who made most of the
  observations under difficult conditions.
  We also thank our referee, E.~Brinks, for careful reading of the
  manuscript and useful comments, which improved the paper.
  This research has made use of the NASA/IPAC Extragalactic Database
  (NED) which is operated by the Jet Propulsion Laboratory, California
  Institute of Technology, under contract with the National
  Aeronautics and Space Administration.
  PMW acknowledges partial support from DFG grant FR 916/6-2.
\end{acknowledgements}

\bibliography{PmW}

\end{document}